\RequirePackage[columnwise, mathlines]{lineno}
\documentclass[aps,prl,superscriptaddress,twocolumn]{revtex4}

\usepackage[colorlinks,citecolor=red,urlcolor=blue,bookmarks=false,hypertexnames=true]{hyperref} 
\usepackage{rotating} 
\usepackage{times}
\usepackage{graphicx,wrapfig}
\usepackage{setspace}
\usepackage{amsmath}
\usepackage{epstopdf}
\usepackage[obeyFinal]{easy-todo}
\usepackage{csquotes}
\usepackage{xr}
\usepackage{etoolbox}

\newcommand*\linenomathpatchAMS[1]{%
  \cspreto{#1}{\linenomathAMS}%
  \cspreto{#1*}{\linenomathAMS}%
  \csappto{end#1}{\endlinenomath}%
  \csappto{end#1*}{\endlinenomath}%
}

\newcommand*\linenomathpatch[1]{%
  \cspreto{#1}{\linenomath}%
  \cspreto{#1*}{\linenomath}%
  \csappto{end#1}{\endlinenomath}%
  \csappto{end#1*}{\endlinenomath}%
}

\expandafter\ifx\linenomath\linenomathWithnumbers
  \let\linenomathAMS\linenomathWithnumbers
  \patchcmd\linenomathAMS{\advance\postdisplaypenalty\linenopenalty}{}{}{}
\else
  \let\linenomathAMS\linenomathNonumbers
\fi

\linenomathpatch{equation}
\linenomathpatchAMS{gather}
\linenomathpatchAMS{multline}
\linenomathpatchAMS{align}
\linenomathpatchAMS{alignat}
\linenomathpatchAMS{flalign}

\begin{document}


\title{Decoupling of dipolar and hydrophobic motions in biological membranes} 

\author{Hanne S. Antila}
\affiliation{Department of Theory and Bio-Systems, Max Planck Institute of Colloids and Interfaces, 14424 Potsdam, Germany}
\author{Anika Wurl}
\affiliation{Institute for Physics, Martin-Luther University Halle-Wittenberg, 06120 Halle (Saale), Germany}
\author{O. H. Samuli Ollila}
\affiliation{Institute of Biotechnology, University of Helsinki, 00014 Helsinki, Finland}
\author{Markus S. Miettinen}
\affiliation{Department of Theory and Bio-Systems, Max Planck Institute of Colloids and Interfaces, 14424 Potsdam, Germany}
\author{Tiago M. Ferreira}
\email[]{tiago.ferreira@physik.uni-halle.de}
\affiliation{Institute for Physics, Martin-Luther University Halle-Wittenberg}



\date{\today}

\begin{abstract}
  Cells use homeostatic mechanisms to maintain an optimal composition of distinct types of phospholipids in cellular membranes. The hydrophilic dipolar layer at the membrane interface, composed of phospholipid headgroups, regulates the interactions between cell membranes and incoming molecules, nanoparticles, and viruses. On the other hand, the membrane hydrophobic core determines membrane thickness and forms an environment for membrane-bound molecules such as transmembrane proteins. A fundamental open question is to what extent the motions of these regions are coupled and, consequently, how strongly the interactions of lipid headgroups with other molecules depend on the properties and composition of the membrane hydrophobic core. We combine advanced solid-state nuclear magnetic resonance spectroscopy methodology with high-fidelity molecular dynamics simulations to demonstrate how the rotational dynamics of choline headgroups remain nearly unchanged (slightly faster) with incorporation of cholesterol into a phospholipid membrane, contrasting the well known extreme slowdown of the other phospholipid segments. Notably, our results suggest a new paradigm where phospholipid headgroups interact as quasi-freely rotating flexible dipoles at the interface, independent of the properties in the hydrophobic region.


\end{abstract}

\maketitle 


\section{Introduction}
Out of the plethora of lipids found in nature, the most ubiquitous are glycerophospholipids which consist of a glycerol backbone attached to two hydrophobic fatty acid chains and a phosphodiester bridge connecting to a hydrophilic headgroup~\cite{Marsh:HLB2}.
Cells use vast amounts of energy, and rely on complex synthetic pathways, to adjust and maintain the specific composition of different types of phospholipid headgroups across cellular organelles~\cite{Har18:281}. This chemical homeostasis implies that the headgroups play a key role in fundamental biological processes, and evidence for phospholipid-specific functionality concerning compartmentalization, signaling, transport, ion binding, peptide insertion, and regulation of membrane protein function has been found~\cite{Lem08:99,Mee08:112,Har18:281,Cor19:5775}. 
However, the molecular details on how lipid composition of a cellular membrane connects to its overall properties and to specific biological processes remain poorly understood.

A key open question is to what degree the behavior of the water-facing headgroups in biological membranes correlate with the properties of the acyl chains in the membrane hydrophobic core~\cite{Raj81:1160,Huang99,Rob04:17066,Ali07:5372,Kla08:3074,Siv09:3420,Moh10:7516,Lee18:2179}. 
Two limiting cases can be considered: 1) the conformational ensemble and dynamics of the headgroup, though positionally connected through the glycerol backbone, are uncoupled from the acyl chain region (freely rotating/weak coupling limit), or 2) the orientation and dynamics of the headgroup and hydrophobic acyl chains are strongly interdependent (strong coupling limit). These two limiting scenarios will give rise to very distinct biophysical behaviour. In the case of strong coupling, the headgroups in lipid domains with ordered acyl chains and hindered dynamics, such as cholesterol-induced lipid rafts~\cite{Sim97:569,Sim00:1721}, would exhibit slower dynamics and possibly a different conformational ensemble. The acyl chain behavior would then indirectly affect the interactions between lipid headgroups and molecules in the aqueous media or within the membrane, such as proteins or drugs. 
Such scenario is implicit, for example, in the popular {\it{umbrella model}} for lipid--cholesterol interactions~\cite{Mee08:112,Huang99,Ali07:5372}. 
In the weak coupling limit, the behavior of lipid headgroups is similar irrespective of the acyl chain structure, order, and dynamics, and consequently any cellular processes that depend on the conformation and dynamics of the headgroups are unaffected by the properties of the hydrophobic region.

We address the question of headgroup-tail decoupling using a
 phosphatidylcholine (PC)--cholesterol bilayer system (the most abundant phospholipid and sterol in eukaryotic cells)---a model cellular membrane from which a wealth of both experimental and simulation data can be obtained. Cholesterol is known to drive lateral heterogeneity and make membranes more ordered (the so-called cholesterol condensing effect), which manifests as a substantial increase in hydrophobic acyl chain C--H bond order parameters ($S_{\rm{CH}}$) in nuclear magnetic resonance (NMR) experiments~\cite{Vis90:415,Hof03:2192,Tib04:132,War05:987,Ver07:919,Dav09:521,Lef11:818,Fer13:1976,And17:E3592}. In contrast, the headgroup and glycerol backbone order parameters 
are essentially unaffected up to the highest cholesterol concentrations possible to incorporate in PC membranes~\cite{Bro78:381,Fer13:73}. The $\alpha$-carbon order parameter of the choline headgroup, in particular, remains unchanged also upon other bilayer perturbations that significantly affect acyl chain order parameters, such as temperature, acyl chain composition, or membrane phase (Table~\ref{tab:SCH} in the supplementary information and references therein). On the other hand, the choline headgroup orientation (and consequently the headgroup order parameters) is highly sensitive to hydration level~\cite{Ulr94:1441}, hydrostatic pressure~\cite{Bon95:518}, and the inclusion of charges~\cite{Alt84:3913,scherer89,Mac91:3558} or molecular dipoles in the membrane~\cite{Bec91:3923}. Therefore, a picture of a rotationaly decoupled headgroup, whose orientation is independent of the hydrophobic region but can be affected by the environment, emerges.  

Although the C--H bond order parameters contain accurate information on conformational ensembles, they do not convey how fast that ensemble is sampled (conformational dynamics). 
The motional time-scales have been dominantly assessed through spin-lattice relaxation ($R_1$) and spin-lattice relaxation in the rotating frame ($R_{1\rho}$) NMR measurements~\cite{Bro79:5045,Bro83:4325,Mor94:749,Kla08:3074,Fer15:44905,Siv09:3420,Rob09:132,Gue95:1952} which are sensitive to different (limited) time-scales depending on experimental conditions and from which a physically meaningful change in the dynamics (speedup vs. slowdown) can be challenging to interpret without multiple measurements under different magnetic fields. Such measurements demonstrate that the cholesterol-induced order in acyl chains is accompanied by slower rotational dynamics not only of the acyl chains but also of the glycerol backbone segments for which there is only a marginal conformational change~\cite{Siv09:3420,Rob09:132}. 

However, in a crucial contrast, the impact of cholesterol on the headgroup motional time-scales, potentially occurring either {\it{via}} direct interaction or through the observed glycerol backbone slowdown, has remained unclear. 
An observation of increase of headgroup $^{13}$C $R_{1\rho}$ rates of DMPC upon incorporation of cholesterol~\cite{Gue95:1952} suggests that the cholesterol-induced slowdown of tail dynamics propagates to the phospholipid headgroups although neither the statistical significance nor the quantitative interpretation of the $R_{1\rho}$ increase in terms of physically meaningful correlation times was provided. In stark contrast, comparison of $^{13}$C cross polarization (CP) and refocused insensitive nuclear enhanced polarization transfer (rINEPT) intensities suggest that the headgroup motional time-scales remain unchanged even by the addition of 50\% cholesterol~\cite{Fer13:1976}.

Here, we show that the dynamics of the PC headgroup are unaffected by cholesterol, and consequently, that the motion of phospholipid headgroups is decoupled from the hydrophobic region (the freely rotating limit). To this end, we employ our novel NMR methodology~\cite{Fer15:44905} where segmental effective correlation times ($\tau_{\rm{e}}$) are determined from solid-state NMR measurements of $R_{1}$, $R_{1\rho}$ and $S_{\rm{CH}}$. The analysis of $\tau_{\rm{e}}$ values enables us to interpret the relaxation rates in terms of a single, physically meaningful average time-scale for each carbon where lower $\tau_{\rm{e}}$) value denotes slowdown and vise-versa. Additionally, we decipher the origin of the decoupled motion by analysing distinct all-atom (CHARMM36 and Slipids) and united-atom (Berger) MD simulations which provide either realistic uncoupled (CHARMM36 and Slipids) or non-realistic coupled (Berger) headgroup motions. The MD simulation models indicate that the decoupled motion originates from dihedral rotations that are present in all other glycerophospholipid types in addition to PCs. This suggests that biological membranes have independent rotational dynamics of the headgroups from the hydrophobic region, a feature that may be relevant in the machinery of biological cell membrane processes.      

\section{Materials and Methods}

\subsection{Sample Preparation}
\small
1-palmitoyl,2-oleoyl-$sn$-glycero-3-phosphocholine (POPC), cholesterol and chloroform were purchased from Sigma-Aldrich. The samples were prepared by mixing the lipids with chloroform and rapidly evaporating the organic solvent under a nitrogen gas flow to obtain a homogeneous lipid film. Subsequently, the lipid film was dried under vacuum overnight. The film was then hydrated in a 0.5 ml tube by adding 50 \%wt of water and manually mixing with a thin metal rod multiple times alternated by sample centrifugation until a homogeneous mixture was attained. The resulting mixture was then centrifuged into a KEL-F Bruker insert with a sample volume of approximately 25 $\mu$l specifically designed for solid-state NMR 4\,mm rotors and left to equilibrate for at least 24 hours at room temperature before measurements. 

\subsection{NMR Experiments}

The solid-state NMR experiments to measure $R_1$ and $R_{1\rho}$ were performed on a Bruker Avance II-500 NMR spectrometer operating at a $^{13}$C Larmor frequency of 125.78 MHz equipped with an E-free CP-MAS 4 mm ($^{13}$C/$^{31}$P/$^1$H). The R-PDLF measurements were performed on a Bruker Avance III 400 spectrometer operating at a $^1$H Larmor frequency of 400.03 MHz equipped with a standard 4 mm CP-MAS HXY probe. All experiments were performed under magic-angle spinning (MAS) conditions at a rate of 5 kHz. The R-PDLF, $R_1$, and $R_{1\rho}$ experiments were performed as previously described in references~\cite{Los18:9751,Fer15:44905}. More details on experimental set up are given in the supplementary information (SI). 



\subsection{MD Simulations}

We performed MD simulations for two systems: a pure POPC bilayer and a bilayer containing additional 50\% cholesterol, both accompanied by enough water molecules per lipid to result in fully hydrated bilayers. We used three lipid MD models (force fields): CHARMM36~\cite{klauda10}, Slipids~\cite{jambeck12b}, and Berger~\cite{ollila07a}/H\"{o}ltje~\cite{holtje01,ferreira13} together with either TIP3P~\cite{Jorgensen:1983a,mackerell98} or SPC~\cite{berendsen81} water. The choice of force fields was based on previous works where their ability to capture structure~\cite{Bot15:15075} and dynamics~\cite{Antila21} of lipid headgroup and glycerol backbone was assessed against NMR measurables. The simulations were performed using the GPU-version of Gromacs2020~\cite{abraham2015gromacs} MD engine, with sampling rate of 10~ps and maintaining 303\,K temperature. The list of all simulated systems, along with the trajectory lengths and links to the freely available simulation data, is given in Table~\ref{tab:sim}. Further details of the simulations are presented in the SI.

\begin{table}[h]
\caption{Summary of simulated systems: the force fields used, numbers of POPC, cholesterol, and water molecules, trajectory lengths, and access links to the simulation files.}
\begin{minipage}[t]{\columnwidth}
\resizebox{\columnwidth}{!} {
\begin{tabular}{ l  c  c  c  c }
\hline
 Force-field POPC/water+cholesterol & POPC/chol & water & length (ns) & files \\

\hline
 CHARMM36~\cite{klauda10}/TIP3P~\cite{mackerell98} & 122/0 & 4480 & 840 & {[}\citenum{Hanne_c36_chol_n_nochol}{]}  \\
   
     Slipids~\cite{jambeck12b}/TIP3P~\cite{Jorgensen:1983a} & 122/0 & 4480 &1200 & {[}\citenum{Hanne_slipids}{]} \\
     
 Berger-POPC-07\cite{ollila07a}/SPC~\cite{berendsen81} & 256/0 & 10342  & 1200 &  {[}\citenum{Hanne_berger_chol}{]}  \\
 
 CHARMM36~\cite{klauda10}/TIP3P~\cite{mackerell98}+CHARMM36~\cite{lim12}  & 122/122 & 9760 & 1240 &  {[}\citenum{Hanne_c36_chol_n_nochol}{]}   \\

 Slipids~\cite{jambeck12b}/TIP3P~\cite{Jorgensen:1983a}+Slipids~\cite{jambeck13chol} & 122/122 & 9760 & 1200 & {[}\citenum{Hanne_slipids_chol}{]} \\
   
Berger-POPC-07~\cite{ollila07a}/SPC~\cite{berendsen81}+H\"{o}ltje-CHOL-13~\cite{holtje01,ferreira13}  & 256/256 & 20480 &1200 & {[}\citenum{Hanne_berger_chol}{]}  \\
\hline

\end{tabular}
}
\end{minipage}
\label{tab:sim}
\end{table}

\normalsize
\begin{figure*}[ht]
  \centering
  \includegraphics[width=16.0cm]{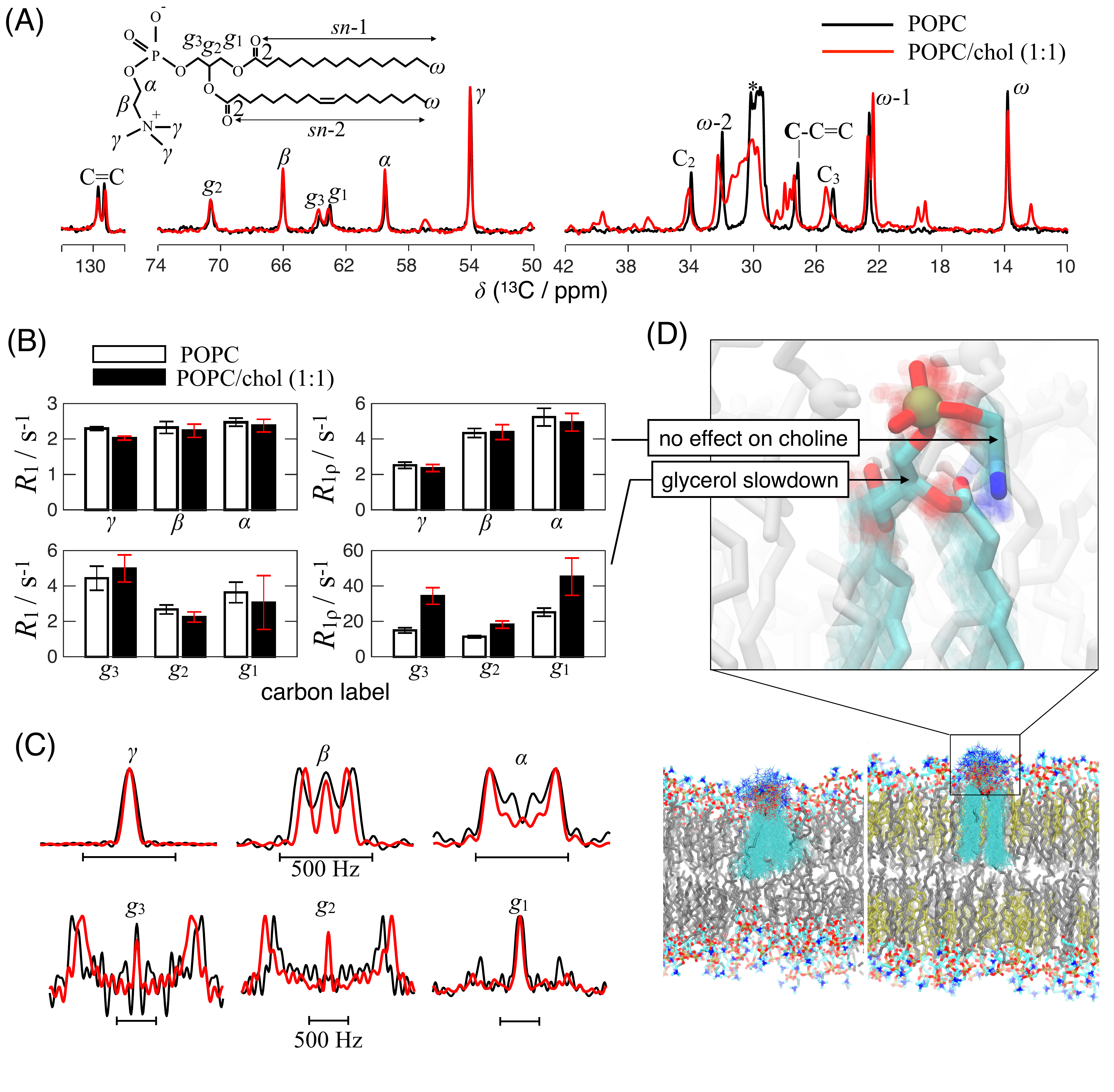}
  \caption{Effect of cholesterol on the dynamics (panel B) and structure (panel C) of headgroup ($\alpha$, $\beta$ and $\gamma$) and glycerol backbone ($g_1$, $g_2$, and $g_3$) carbons in POPC lipid membranes. (A) Chemical structure of POPC with carbon labels and refocused-INEPT spectra from POPC membranes with (red) and without (black) cholesterol. (B) $^{13}$C spin-lattice relaxation rates, $R_1$, and spin-lattice relaxation in the rotating frame rates, $R_{1\rho}$, showing the independent motion of the headgroup and slowdown of the glycerol backbone upon cholesterol incorporation. A Larmor frequency of 500 MHz for $^1$H nuclei and a spin lock field equal to 50 kHz were used. The corresponding experimental decays for each data value are shown in Figure~\ref{SI_R1_R1rho}. (C) Dipolar recoupling profiles acquired with R-PDLF spectroscopy from POPC membranes with (red) and without (black) cholesterol. Similar magnitudes of the splittings indicate structural independence of both the headgroup and glycerol backbone on cholesterol incorporation. Note that the line shapes are highly sensitive to the experimental setup and that the relevant information on conformations is in the splittings, which are proportional to $|S_{\rm{CH}}|$. (D) Overlayed lipid conformations in MD simulations with ({\it right}) and without ({\it left}) cholesterol (yellow) illustrating the experimental observations. }\label{ExpResults}
\end{figure*} 

\section{Results}

\section{Experimental demonstration of the uncoupled motion}

Figure~\ref{ExpResults} shows the effect of cholesterol on both the dynamics and structure of POPC headgroup and glycerol backbone. Chemical shift resolution for all the distinct carbons in the refocused-INEPT spectra displayed in Figure~\ref{ExpResults}A is enabled by simultaneous magic-angle spinning and heteronuclear decoupling. The effect of cholesterol on the phospholipid dynamics is assessed by measuring the $R_1$ and $R_{1\rho}$ values from POPC and POPC/cholesterol (1:1) multi-lamellar vesicles and the effect on phospholipid structure by R-PDLF spectroscopy. The relaxation rates for the headgroup and glycerol backbone are shown in Figure~\ref{ExpResults}B. The complete set of decays and relaxation rates measured for all the phospholipid segments resolved in the $^{13}$C spectrum is given in supplementary information Figures~\ref{SI_R1_R1rho} (headgroup and glycerol backbone) and \ref{R1_R1rho_tails} (acyl chains). The dipolar splittings used to calculate the $S_{\rm{CH}}$ order parameters of headgroup and glycerol backbone are presented in Figure~\ref{ExpResults}C. The resulting order parameters confirm the previously reported values~\cite{Fer13:1976}.  

$R_1$ rates remain constant for both the glycerol backbone and the headgroup, showing that the C--H bond motions with time-scales close to ns are not affected by cholesterol for these carbons. On the other hand, the $R_{1\rho}$ rates in the glycerol backbone (carbons $g_1$, $g_2$, and $g_3$)
increase by approximately a factor of two upon cholesterol addition. In sharp contrast, the $R_{1\rho}$ values for the choline headgroup ($\alpha$, $\beta$, and $\gamma$ segments) are unaffected by cholesterol and significantly lower than in the glycerol backbone. The invariance of both the headgroup carbon dipolar couplings (Figure~\ref{ExpResults}C) and the relaxation rates (Figure~\ref{ExpResults}B) upon cholesterol incorporation implies that the conformational ensemble and the time required to span all the available conformations is the same irrespective of the glycerol backbone slowdown induced by cholesterol. For acyl chains both structural ($S_{\rm{CH}}$) and dynamic observables ($R_1$ and $R_{1\rho}$) vary with incorporation of cholesterol (SI Figures \ref{R1_R1rho_tails} and \ref{fig_SCH_Exp_CHARMM}), signaling the expected ordering and slowdown in line previously reported results~\cite{Fer13:1976,Siv09:3420}.

\begin{figure}
  \centering
  \includegraphics[width=9cm]{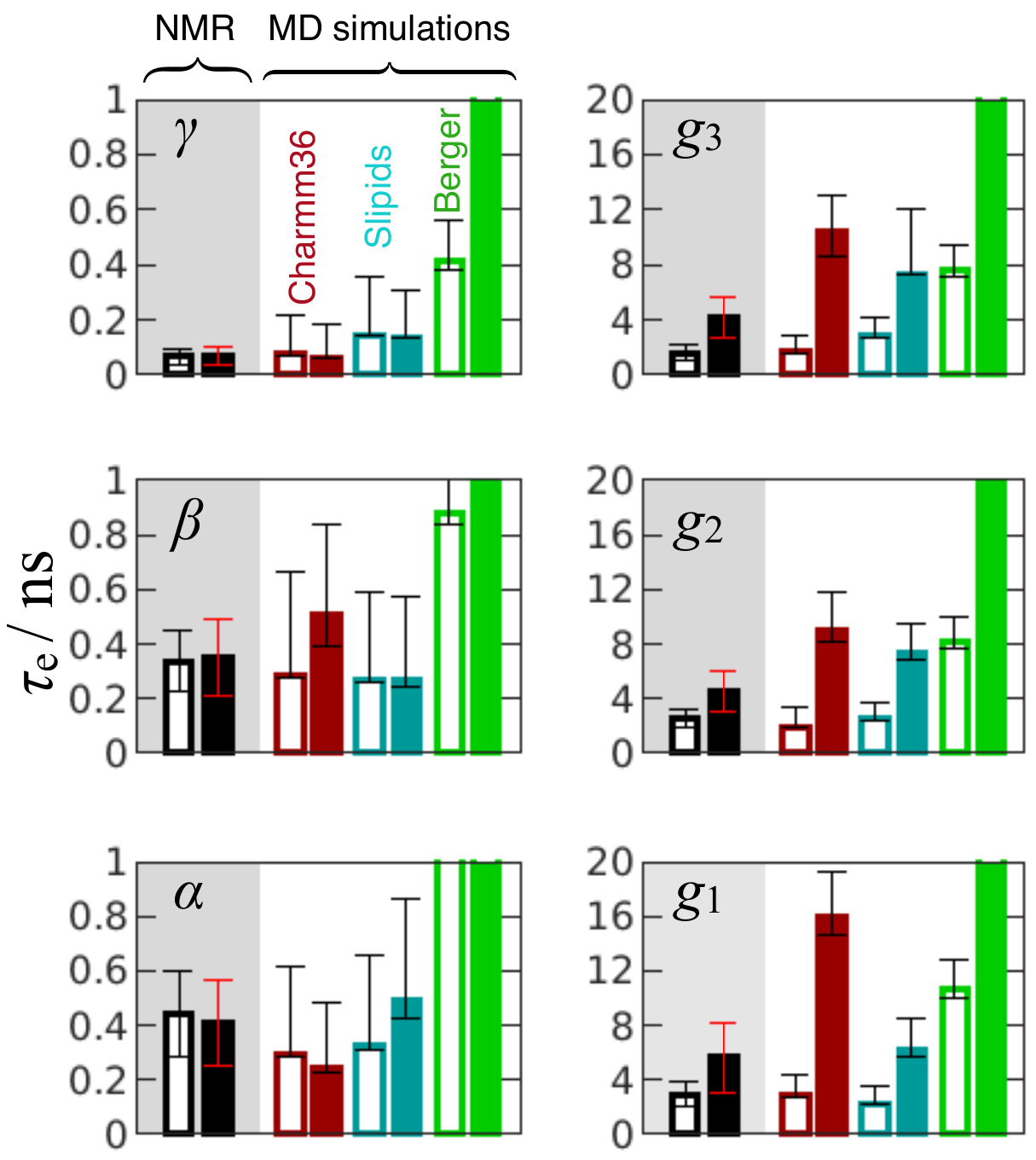}
  \caption{Impact of cholesterol (hollow bars: pure POPC, filled bars: 50\% POPC+50\% cholesterol) on the effective correlation times, $\tau_{\rm{e}}$, of different carbons in the headgroup and glycerol backbone of POPC quantified experimentally and from lipid bilayer MD simulations with the CHARMM36, Slipids and Berger force-fields. Note the different $y$-scales used on the left and right plots to appreciate the significant difference of effective correlation times for the choline headgroup (0.1-0.5 ns) and glycerol backbone segments (2-5 ns). }\label{fig_taue}
\end{figure}  

To give an intuitive measure, sensitive to all time-scales, for the headgroup and glycerol backbone dynamics, we quantified the effective correlation times $\tau_e$. (Figure~\ref{fig_taue}). To this end, we used the experimental data presented in panels B and C of Figure~\ref{ExpResults} to calculate~\cite{Fer15:44905}
\begin{equation}
\tau_e=\frac{5R_{1\rho}-3.82R_1}{4\pi^2 d_{\rm{CH}}^2N(1-S^2_{\rm{CH}})},
\label{tau_eff_eq}
\end{equation}
where $N$ denotes the number of protons covalently bound to the carbon and the coupling constant $d_{\rm{CH}}$ is approximately -22~kHz. The $\tau_e$ values of the choline headgroup are within 0.1--0.5~ns and remain constant within the experimental accuracy upon addition of 50\% cholesterol, while the
$\tau_{\rm{e}}$ values for 
the glycerol backbone slowdown with a factor of approximately two, and are one order of magnitude slower than in the headgroup. 
\section{Comparison of experiments with MD simulations}

\begin{figure}
  \centering
  \includegraphics[width=8.3cm]{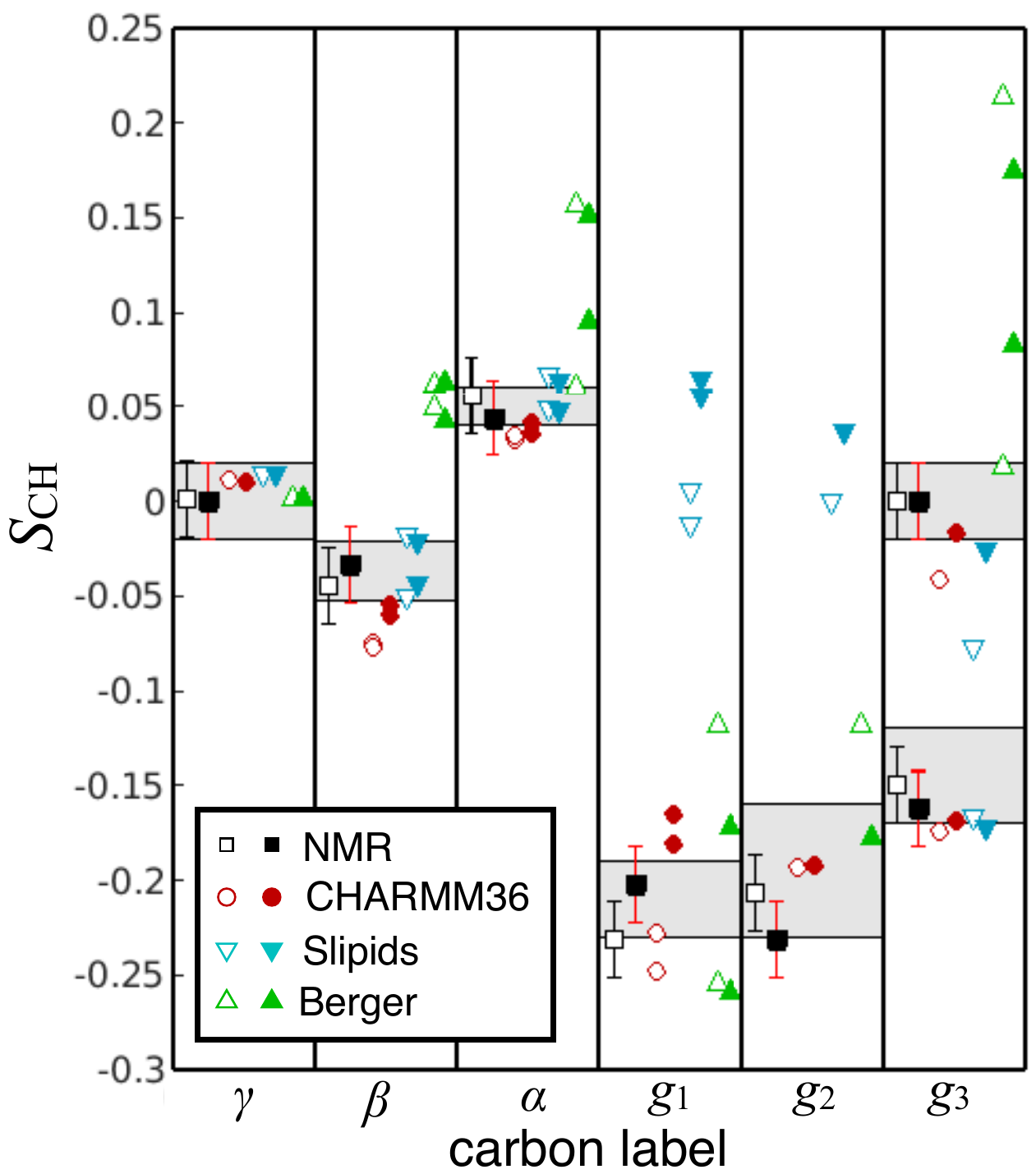}
  \caption{Effect of cholesterol (hollow symbols: pure POPC, filled symbols: 50\% POPC+50\% cholesterol) on the  POPC headgroup and glycerol backbone C--H bond order parameters. The experimental values were determined by R-PDLF spectroscopy. The grey areas show the range of $S_{\rm{CH}}$ values for PC bilayer systems with and without cholesterol reported until date (see e.g. References~\cite{Lef11:818,Fer13:1976}). For comparing with the effect on the acyl chains see Figure~\ref{fig_SCH_Exp_CHARMM}.  }\label{fig_SCH}
\end{figure}

\begin{figure*}[ht]
  \centering
  \includegraphics[width=16cm]{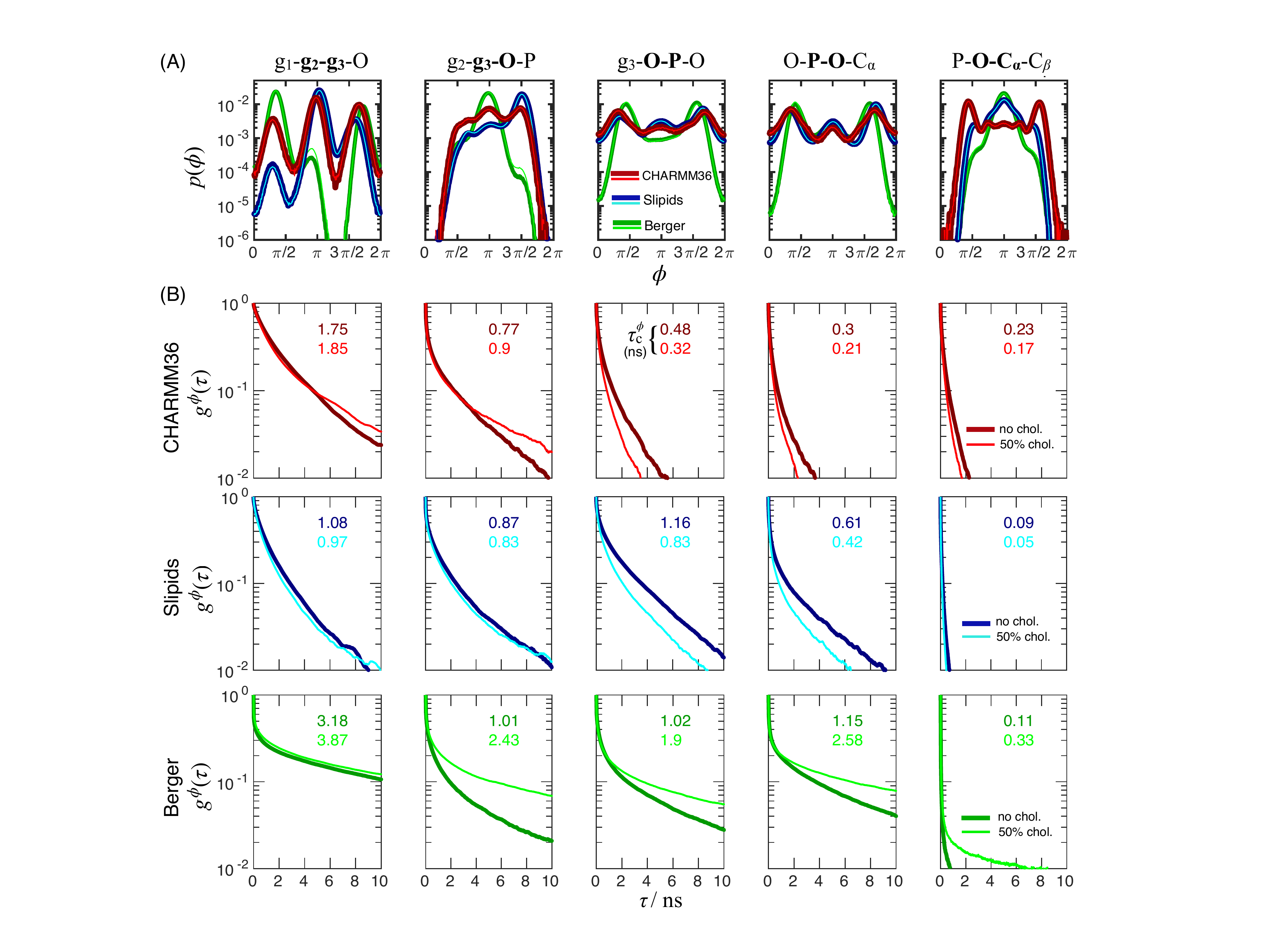}

  \caption{Effect of cholesterol on internal structure and dynamics of POPC. (A) Dihedral angle distributions for pure POPC membranes (darker thick lines) and POPC/cholesterol membranes (lighter thin lines) from MD simulations using the CHARMM36 (red), Slipids (blue) and Berger (green) force-fields. (B) Reduced and normalized dihedral torsion autocorrelation functions (see Eq.~\ref{rednormdih}) showing their corresponding dihedral effective correlation times ($\tau_c^\phi$). Note that the extremely short correlation times for the outermost right column are due to the very narrow angle range accessible for this torsion dihedral.    }\label{fig_dihedrals}
\end{figure*}

Also included in Figure~\ref{fig_taue} are the $\tau_{\rm{e}}$ values calculated from three sets of MD simulations using the CHARMM36, Slipids, and Berger force-fields. 
Notably, CHARMM36 simulations reproduce the experimental $\tau_{\rm{e}}$ values of the choline headgroup, both with and without cholesterol, flawlessly within experimental uncertainty. CHARMM36 also captures well the experimental choline C-H bond order parameters which remain the same with the presence of the sterol (Figure~\ref{fig_SCH}). For the glycerol backbone, CHARMM36 simulations slightly overestimate the slowdown of the effective correlation times, but give the best structural model among the three force-fields used.
Slipids simulations show the best agreement with experiments for the effect of cholesterol on the glycerol backbone $\tau_{\rm{e}}$ values, although they fail to capture the glycerol backbone structure, and overestimate the effective correlation time for the $\gamma$ carbon. 

The Berger force-field clearly produces the least realistic dynamics, giving a significant overestimation of $\tau_{\rm{e}}$ for the choline headgroup segments (both with and without cholesterol) and predicts an erroneous, large (approximately 5-fold) cholesterol-induced slowdown of the choline headgroup dynamics (see Figure~\ref{fig_taue_Berger} for $\tau_e$s from the Berger simulations drawn to scale). Both the structural and dynamical force field properties observed here are in line with the previously reported~\cite{Bot15:15075,Antila21}. 

In contrast to the Berger model, both the CHARMM36 and Slipids models, although not perfect, capture the key experimental observation in this work: The structure and dynamics of the choline headgroup are not affected by incorporation of cholesterol despite the increased acyl chain order and hindered dynamics in the acyl chains and glycerol backbone, i.e, the headgroup is uncoupled from the glycerol backbone in these models.

\section{Effect of cholesterol on the time-scale of internal motions of the headgroup and glycerol backbone}

Knowing their ability to reproduce the NMR measurables, we proceed to exploit the temporal and spatial resolution in the distinct MD models to gain insight on the rotational dynamics of specific sites on the molecules as well as the origin and degree of (un)coupling between the headgroup and the rest of the phospholipid.  
    To this end, Figure~\ref{fig_dihedrals} shows the distributions of selected headgroup and glycerol dihedral angles, $\phi$, and the corresponding dihedral effective correlation times, $\tau_c^\phi$, which are extracted from autocorrelation functions 
    \begin{equation}
      G^\phi(\tau)=\langle \cos\phi(t)\cos\phi(t+\tau)\rangle, 
      \label{gdih}
    \end{equation}
    where the angular brackets denote an average over time and over the number of molecules in the system.   
    We define the dihedral effective correlation time ($\tau_c^\phi$) as simply the area under the reduced normalised autocorrelation function 
        \begin{equation}
          g^\phi(\tau)=\frac{G^\phi(\tau)/G^\phi(0)- \langle \cos\phi\rangle^2}{1- \langle \cos\phi\rangle^2}, \label{rednormdih}
    \end{equation}
    where $\langle \cos\phi\rangle^2$ is the value of $G^\phi(\tau)$ at infinitely long $\tau$. $\tau_c^\phi$ provides a measure of how much time is needed for dihedral motion to sample its angle distribution. 
    
    For the more realistic force-fields (CHARMM36 and Slipids) cholesterol does not affect the dihedral distributions of POPC as expected from the $S_{\rm{CH}}$ data alone. More interestingly, for both of these force-fields, the dihedral angles over the phosphate linkage between the glycerol and the alpha carbon, ($g_3$)O--P(O) and (O)P--O($\alpha$), and the glycerol backbone dihedral $g_2$--$g_3$, have no angles which are unavailable, in contrast to the other dihedrals analysed. Furthermore, the ($g_3$)O--P(O) and (O)P--O($\alpha$) exhibit  fast correlation times ($\tau_c^\phi\leq$0.5~ns) which are very close to the effective correlation times in Figure~\ref{fig_taue}.  
    
    The most notable differences between the more realistic simulations and Berger simulations are the slower $\tau_c^\phi$ in Berger, and how these are affected by cholesterol. While in CHARMM36 and Slipids only minor changes are observed, with a slight speedup of the sampling by cholesterol, in the Berger model these internal dynamics slowdown considerably. 

\section{Polar and azimuthal motion of the choline dipole and of the glycerol backbone}

To investigate the correlation between motions of headgroup and other parts of lipid molecules, we quantified the autocorrelation functions of the polar and azimuthal angles, $\theta$ and $\varphi$ (coordinate system where the z-direction coincides with membrane normal), for a number of selected vectors between intramolecular atomic pairs. The definition of the autocorrelation functions $G^\theta(\tau)$ and $G^\varphi(\tau)$ are the same as in Eq.~\ref{gdih} but using $\theta$ and $\varphi$ as angles, respectively. Note that for $G^\theta(\tau)$ a non-zero plateau at the long $\tau$ is expected since the different $\theta$ angles are not equally likely to occur. On the other hand, $G^\varphi(\tau)$ is always zero at long $\tau$ due to the lipid uniaxial motion. 

\begin{figure}
  \centering
  \includegraphics[width=8.5cm]{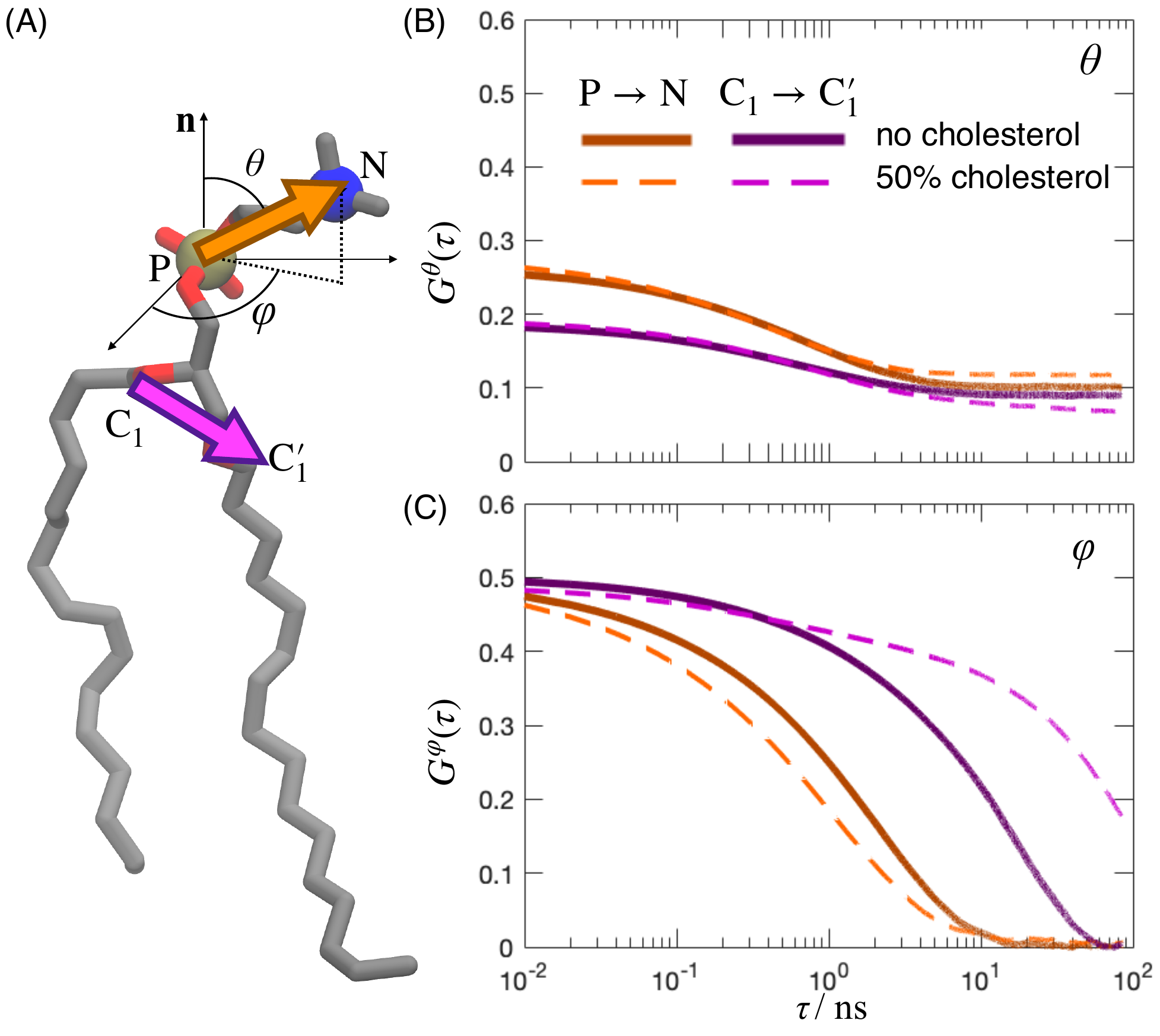}
  \caption{The effect of cholesterol on the time-scales for the reorientations of P$\rightarrow$N and ({{\it{sn}}-1)O=C$\rightarrow$C}=O({\it{sn}}-2) vectors over the spherical angles $\varphi$ and $\theta$. The autocorrelation functions shown here are described in Equation~\ref{gdih}. }  \label{varphi_theta}
  
\end{figure}

In Figure~\ref{varphi_theta} we show these correlation functions for the choline dipole orientation, P$\rightarrow$N, and for the interatomic vector connecting the carbonyl carbons in the {\it sn}-1 and {\it sn}-2 positions, calculated from the most realistic force-field (CHARMM36). The $ \varphi$ autocorrelation functions clearly show the contrasting effects of cholesterol on these vectors. While for the choline dipole a speedup of reorientational motion is observed, the $\varphi$ dynamics of the vector connecting the carbonyl carbons become slower by almost an order of magnitude. To extract effective correlation times $\tau_c^\theta$ and $\tau_c^\varphi$, we again integrate the reduced and normalised autocorrelation functions. While cholesterol induces more than a 4-fold slowdown for the $\varphi$ dynamics of the vector connecting carbonyl carbons, the correlation times of the P$\rightarrow$N orientation remain essentially the same. The complete set of reduced autocorrelations functions analysed is given in SI Figures~\ref{PN_phi_theta}-\ref{C21_C31_phi_theta} together with $\theta$ distributions, and $\tau_c^\theta$ and $\tau_c^\varphi$ values. 

\section{Discussion}

Our experimental and MD simulation results show that the phospholipid headgroup conformational ensemble and dynamics remain unaffected by addition of 50\% cholesterol to the lipid membrane, despite the significant acyl chain ordering and reduction in both the acyl chain and glycerol backbone dynamics. Therefore, our results do not support models that contain interdependence between the structure or dynamics of the hydrophobic and hydrophilic regions of cellular or model phospholipid membranes. 

The observed slowdown of the glycerol backbone upon addition of cholesterol (Figures \ref{fig_taue} and \ref{taue_tails}) arises from the longer timescales
to which $R_{1\rho}$ is sensitive to. The internal motions of the glycerol backbone are not affected by cholesterol in the most realistic MD simulations used (CHARMM36 and Slipids, see Figure~\ref{fig_dihedrals}) in line with the invariance of the glycerol backbone $R_1$ values (Figure~\ref{ExpResults}). Therefore, the slower glycerol backbone dynamics induced by cholesterol most likely arises from a slowdown of the 
rotational diffusion of the whole phospholipid body as previously suggested by Roberts et al.~\cite{Siv09:3420,Rob09:132}, rather than restrictions in internal dynamics.  

The independence of headgroup and hydrophobic chain motions must result from a set of fast internal rotations around phospholipid bonds with specific orientations that decouple these motions. The MD simulations in best agreement with the NMR experiments show a high flexibility for dihedral angles in the headgroup region with a wide range of accessible conformations (Figures~\ref{fig_dihedrals},~\ref{fig_dihedral_non_log_scale} as well as Ref.~\cite{bacle21}). From the set of dihedral distribution functions, one clearly observes the highly flexible nature of the ($g_3$)O--P,  P--O($\alpha$) and $g_2$-$g_3$ dihedral angles with all angles over a complete dihedral rotation having non-zero probability in contrast to the remaining torsions. This applies for all three force-fields, though Berger has a less even distribution than CHARMM36 and Slipids.  
For CHARMM36 and Slipids, the correlation times for the rotations around the ($g_3$)O--P and P--O($\alpha$) bonds are lower than 0.5\,ns (Figure~\ref{fig_dihedrals}) and very close to the effective correlation times measured for the C--H bonds from the $\alpha$ and $\beta$ carbons. These $\tau_c^\phi$ values slightly decrease with the addition of cholesterol, i.e. cholesterol induces a slight speed-up of the torsion dynamics for these particular dihedrals, most likely because fewer steric hindrances are present due to an increased average distance between headgroups. 

The highly flexible dihedral rotations around the phosphate P--O bonds, as well as the g2--g3 torsion, are much faster than the transverse and longitudinal rotational diffusion of the molecular frame whose correlation times have been approximated to be 10--20~ns and 100~ns, respectively~\cite{Kla08:3074}. The most probable $\theta$ angle of the the (g3--)O--P and g2--g3 bonds is less than 10$^\circ$ (Figures~\ref{O11_P_phi_theta} and \ref{C2_C1_phi_theta}) which is very close to be parallel with the bilayer normal axis. A flexible, fast-rotating dihedral aligned with the membrane normal leads to decoupling of headgroup from the longitudinal rotational diffusion of the molecular frame. Decoupling from the transverse rotational diffusion is mostly enabled by the fast motion over the P--O(--C$_\alpha$).

The torsions around the phosphate group and g2--g3 dihedrals act analogously to a {\it{frictionless spherical-joint}} which decouples the choline headgroup structure and dynamics from the glycerol backbone.
This is in line with the previous analysis~\cite{Kla08:3074} where a partial decoupling of the headgroup from the main phospholipid body due to the rotation of a phosphate dihedral was suggested based on a comparison of CHARMM C27r MD simulation of pure dipalmitoylphosphatidylcholine bilayers to $^{31}$P-NMR R$_{1}$ data under several magnetic fields. Here, we demonstrate that such decoupling is strong enough to prevent the propagation of the slowdown effect of cholesterol from the acyl chains and glycerol backbone to the headgroup. 

We base our molecular interpretation of the decoupled motion on the CHARMM36 force field, which gives the best overall description for the headgroup and glycerol backbone structure and dynamics among the available models (Figures~\ref{fig_taue} and ~\ref{fig_SCH}, and Refs.~\cite{Bot15:15075,Antila21}). However, not all the NMR observables calculated are within experimental errors even in CHARMM36 simulations and we cannot exclude the possibility that improved future models correctly capturing the decoupling effect, effective correlation times, and structural order parameters may give an alternative molecular interpretation. 

The headgroup decoupling is not observed in Berger force-field, although it also provides fast dynamics over the phosphate group dihedrals (Figure~\ref{fig_dihedrals}). This is most likely due to an overestimation of the attractive interaction between cholesterol and the choline group. Such interaction has been interpreted previously as a consequence of the so-called {\it{umbrella effect}} where a reorientation of the headgroup due to presence of cholesterol is often assumed~\cite{alwarawrah12}. However, the combination of MD simulations and experiments presented here indicates that such interaction is artefactual and that both the orientation and dynamics of the headgroup are unaffected by the sterol presence. The implicit assumption in the {\it{umbrella model}} of a cholesterol effect on headgroup reorientation, either through a change of the conformational ensemble or a change of dynamics, is not supported by our experimental results or the more realistic MD models.  
 
Correlation functions of P$\rightarrow$N and other intramolecular vectors, calculated from CHARMM36 simulations, further support the idea of decoupled motions between headgroup and the acyl chains.
Cholesterol induces a significant slowdown of the reorientation of 
the interatomic vectors between atoms belonging to the glycerol backbone and acyl chain segments 
(Figures~\ref{PN_phi_theta}-\ref{C21_C31_phi_theta}), while the effect on
the P$\rightarrow$N vector, representing the choline dipole, is negligible with only a slight speed up of the dynamics most likely due to the increase of the distance between phospholipids headgroups.  

The molecular description suggested here has rather strong implications for membrane biophysics and should motivate a number of additional experiments and simulations.
It implies that the dipolar surface of glycerophospholipid bilayers consists of freely rotating dipoles with timescales faster than 2~ns that do not depend on the dynamics of the acyl chains or glycerol backbone. The time-scale of reorientation of the dipoles is expected to influence the interaction of the headgroups with charged molecules, e.g. proteins, that approach the lipid biomembrane.
For instance, it is known that the tilt of the headgroup dipole is highly sensitive to membrane surface charge~\cite{scherer89}. Under a positive surface charge,  
the headgroups tilt to a more upright orientation (increase of the alpha and beta $S_\mathrm{CH}$ values) and vice-versa for a negative surface charge due to the charge-dipole electrostatic interactions.  The results presented here suggest that the phosphate and the g2--g3 dihedrals enable an unconstrained response of headgroups to the electrostatic field and effectively uncouple the interactions occurring in the membrane surface from the hydrophobic region. Although we only investigate here PC headgroups, it is foreseable that the decoupling applies to all other glycerophospholipids since the same {\it{ molecular bearings}} are present irrespectively of the substituent headgroup~\cite{Marsh:HLB2}. 


In summary, our results suggest that for describing the dipolar interactions at the surface of membranes, the hydrophobic structure may be neglected to a good approximation and that the relevant headgroup physics lie on the electrostatic interactions---which is remarkably useful considering the complex molecular arrangement in the hydrophobic region of biological membranes.  \\

\section{Acknowledgments}

O.H.S.O acknowledges CSC -- IT Center for Science for computational resources and Academy of Finland (grants 315596 and 319902) for financial support.
H.S.A. gratefully acknowledges financial support from the Osk. Huttunen Foundation, Finnish
Academy of Science and Letters (Foundations’ Post Doc Pool), Instrumentarium Science Foundation, and the Alexander von Humboldt Foundation. T.M.F. was supported by the Ministry of Economics, Science
and Digitalisation of the State of Saxony-Anhalt, Germany.\\

T.M.F. greatly acknowledges Kay Saalw{\"a}chter and Alexey Krushelnitsky for invaluable support and discussions. 

\section{Competing interests}
The authors declare no competing interests.

\bibliography{refs.bib,simulation.bib}

\clearpage

\section{Supplementary Information}

\renewcommand{\thefigure}{S\arabic{figure}}
\setcounter{figure}{0}

\renewcommand{\thetable}{S\arabic{table}}
\setcounter{table}{0}

\section{Solid-state NMR experiments}


\subsection{R-PDLF experiments} 
A total of 32 points in the indirect dimension with increments equal to two R18 blocks; SPINAL64 was used for proton decoupling during $^{13}$C acquisition, with a nutation frequency of approximately 50 kHz, a total acquisition time of 0.07 s and a spectral width of 200 ppm; the rINEPT pulses were set at a nutation frequency of 78.12 kHz. \\

\subsection{$R_1$ and $R_{1\rho}$ experiments} 

RF $\pi/2$ and $\pi$ pulses were set to a nutation frequency of 63.45 kHz. TPPM was used for proton decoupling during $^{13}$C acquisition, with a nutation frequency of approximately 50 kHz, a total acquisition time of 0.1 s, recycle delay of 10 s and a spectral width of 140 ppm. The spin-lock frequency for $R_{1\rho}$ was 50 kHz. 

For quantifying $R_1$ and $R_{1\rho}$ for a given carbon segment, we determined the decay over the indirect dimension by fitting gaussian lineshapes in the direct dimension and using the analytic areas of the fitted functions. The decay was then fitted with a single exponential decay and the error bounds for both the $R_1$ and $R_{1\rho}$ values presented are the 95 \% confidence bounds from these fits.

\section{MD simulations}
The force field parameters were acquired from CHARMM-GUI (CHARMM36), the Slipids web page (\url{http://www.fos.su.se/~sasha/SLipids/}, the pre 2020 version), and from existing simulations in zenodo repository~\cite{bergerFILESpopc50chol,bergerFILESpopcT300} (Berger). Similarly, the initial configurations, where the bilayers spanned the x--y plane of a rectangular simulation box with periodic boundary condition, were obtained either from CHARMM-GUI or constructed from existing trajectories~\cite{bergerFILESpopc50chol,bergerFILESpopcT300}. The initial configurations were then relaxed using the steepest decent algorithm (5000 steps), followed by 2.5~ns + 41~ns of NPT simulation with semi-isotropic pressure coupling maintaining 1~bar pressure. In the first 2.5~ns the pressure set using the Berendsen barostat~\cite{berendsen84} with coupling constant 1~ps while the rest of the 41~ns pre-equilibration was done by utilizing the Parrinello--Rahman barostat~\cite{parrinello81} with 1~ps coupling constant. The latter barostat settings where then used for the following 0.84~$\mu$s-1.24~$\mu$s of NPT production runs used for analysis (See Table~\ref{tab:sim} in main text). The temperature in the simulations was set to 303~K using the Nose-Hoover thermostat~\cite{nose84,hoover85} with coupling constant 1~ps. The electrostatic interactions were modelled utilizing the particle-mesh-Ewald algorithm~\cite{essman95}. For CHARMM36 and Slipids simulations a cutoff of 1.2~nm was used to separate the real and reciprocal space contributions to electrostatics whereas the Lennard-Jones potentials were smoothly fixed to zero between cutoffs 1~nm and 1.2~nm. In Berger simulations we utilized verlet list scheme with cutoff 1~nm for both electrostatic and Lennard-Jones interactions.
The leap-frog algorithm with time step 0.002~ps was used to integrate the movement of the particles. To avoid the need for a shorter time-step, the fast vibrations of the C--H bonds of the lipids where removed using the LINCS algorithm~\cite{hess08} and SETTLE~\cite{miyamoto92} constrains were used for water.
The resulting simulation trajectories were analysed using in-house scripts. The exact methodology utilized for extracting the NMR measurables, and their error estimates, from the simulation is detailed in Ref.~\citenum{Antila21}.

\begin{table*}[t]

\begin{tabular}{ c  c  c  c  c  c  c  c }
\hline
 System & method & conditions &$T$ / $^\circ$C & Phase & $S^\alpha_{\rm{CH}}$ & $-S^\beta_{\rm{CH}}$ & ref. \\

\hline
\hline

 DMPC & $^2$H NMR & 150 mM NaCl, 10 mM HEPES &30 & L$_\alpha$ & 0.048 & 0.046 & \cite{Mcd91:3558}\\
   \hline

  DMPC & $^2$H NMR &  150 mM NaCl, 10 mM HEPES &15 & L$_\beta$ & 0.051 & 0.062 & \cite{Mcd91:3558} \\
     \hline

  DOPC & $^2$H NMR & water only &30 & L$_\alpha$ & 0.049 & 0.029 &\cite{Ulr94:1441}\\
  \hline
  POPC & $^2$H NMR & NaCl saturated solution & 23 & L$_\alpha$ & 0.048 & 0.044 &\cite{Bec91:1}\\
   \hline
  DPPC & $^2$H NMR & water only&45-90 & L$_\alpha$ & 0.048 & 0.046-0.024 & \cite{Gal75:3647}\\
  \hline
  DPPC/chol. (1:1) & $^2$H NMR & 0.2 M sodium acetate/acetic acid &10-70 & L$_\alpha$ & 0.048/0.048-0.040 & 0.03-0.024 &\cite{Bro78:381}\\
   \hline
POPC/chol. (1:1) & R-PDLF & water only &30 & L$_\alpha$ & 0.052 & 0.034 & here\\
\hline
POPC & R-PDLF & water only &30 & L$_\alpha$ & 0.052 & 0.040 & here\\
\hline

\end{tabular}
\caption{Previously published $\alpha$ and $\beta$ C--H bond order parameters from $^2$H NMR spectroscopy for a number of phosphatidylcholine lamellar systems together with values reported here using $^1$H-$^{13}$C dipolar recoupling on POPC and POPC/cholesterol (1:1). }
\label{tab:SCH}
\end{table*}

  \begin{figure*}[h]
  \centering
  \includegraphics[width=18.0cm]{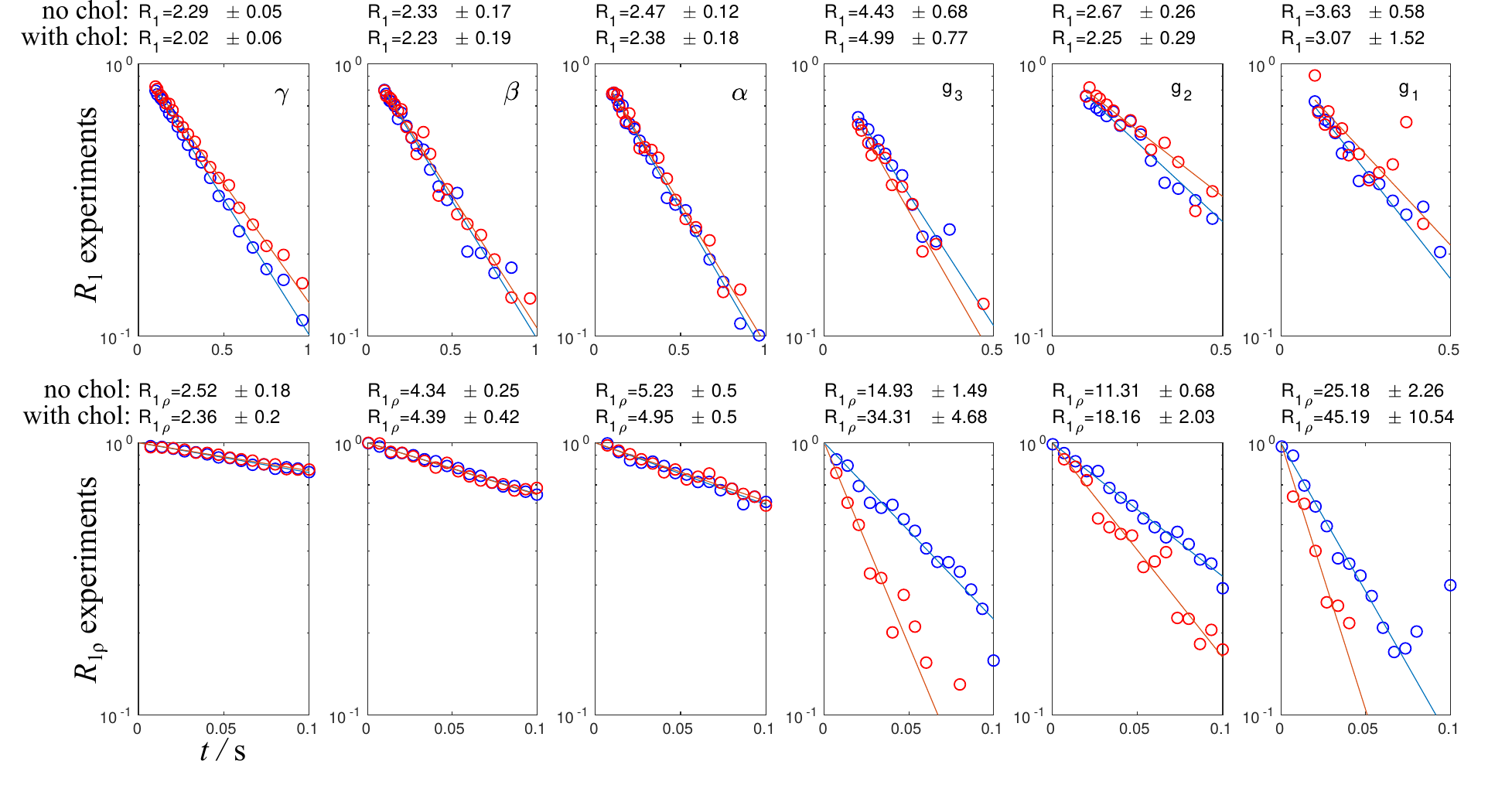}
  \caption{The $R_1$ and $R_{1\rho}$ decays measured for the headgroup and glycerol backbone carbons in the POPC (blue) and POPC/cholesterol (red) systems. Each point corresponds to the integral determined from a Gaussian fit of the corresponding $^{13}$C peak in the high resolution chemical shift spectrum acquired under MAS of 5 kHz. The spin lock field for the $R_{1\rho}$ measurement was 50 kHz and the $^{13}$C Larmor frequency was 125 MHz.}
  \label{SI_R1_R1rho}
\end{figure*}

  \begin{figure*}[h]
  \centering
  \includegraphics[width=18.0cm]{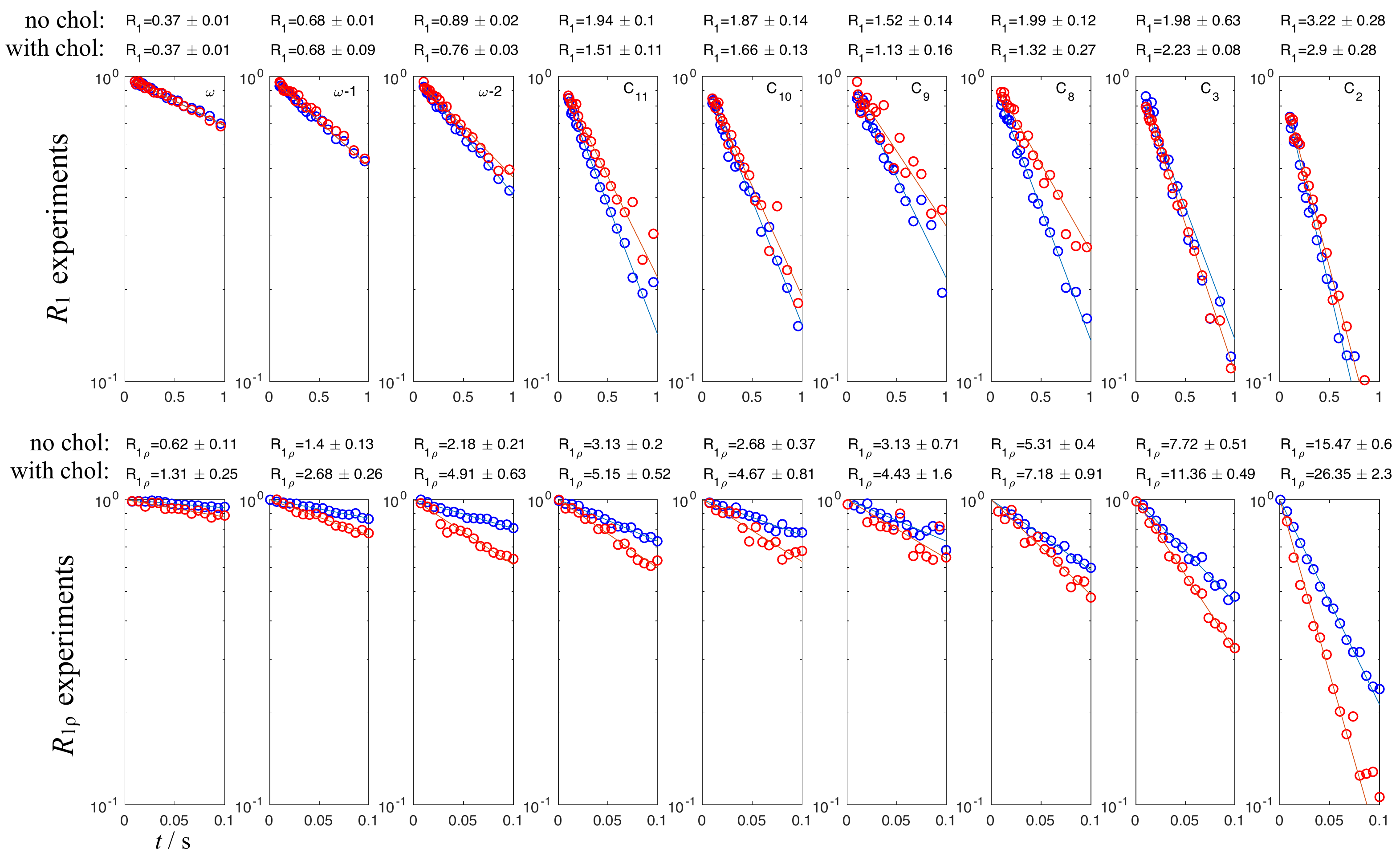}
  \caption{The $R_1$ and $R_{1\rho}$ decays for the acyl chain carbons measured in the POPC (blue) and POPC/cholesterol (red) systems. Each point corresponds to the integral determined from a gaussian fit of the corresponding $^{13}$C peak in the high resolution chemical shift spectrum acquired under MAS of 5 kHz. The spin lock field for the $R_{1\rho}$ measurement was 50 kHz and the $^{13}$C Larmor frequency was 125 MHz.}
 \label{R1_R1rho_tails}
\end{figure*}

\begin{figure*}
  \centering
  \includegraphics[width=19cm]{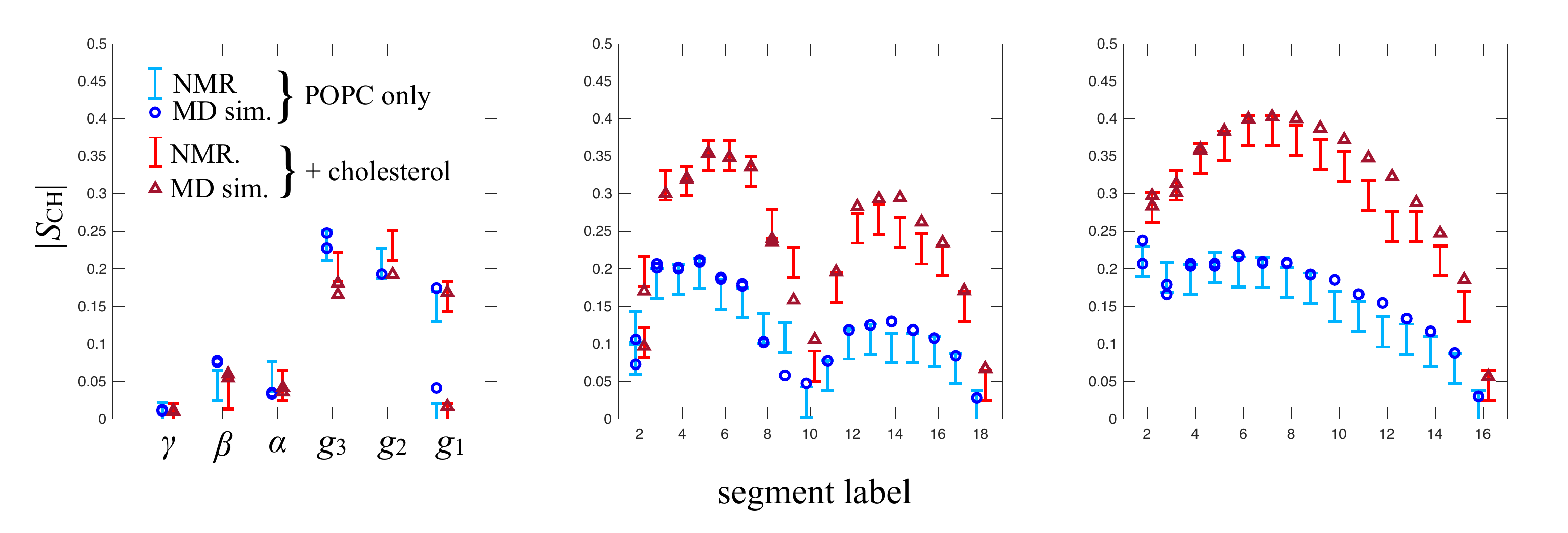}
    \caption{Effect of cholesterol on the C--H bond order parameter magnitudes, $|S_{\rm{CH}}|$, of different segments in the acyl chains of POPC measured experimentally and calculated from lipid bilayer MD simulations with the CHARMM36 force-field.}
\label{fig_SCH_Exp_CHARMM}
\end{figure*}

\begin{figure*}
  \centering
  \includegraphics[width=10cm]{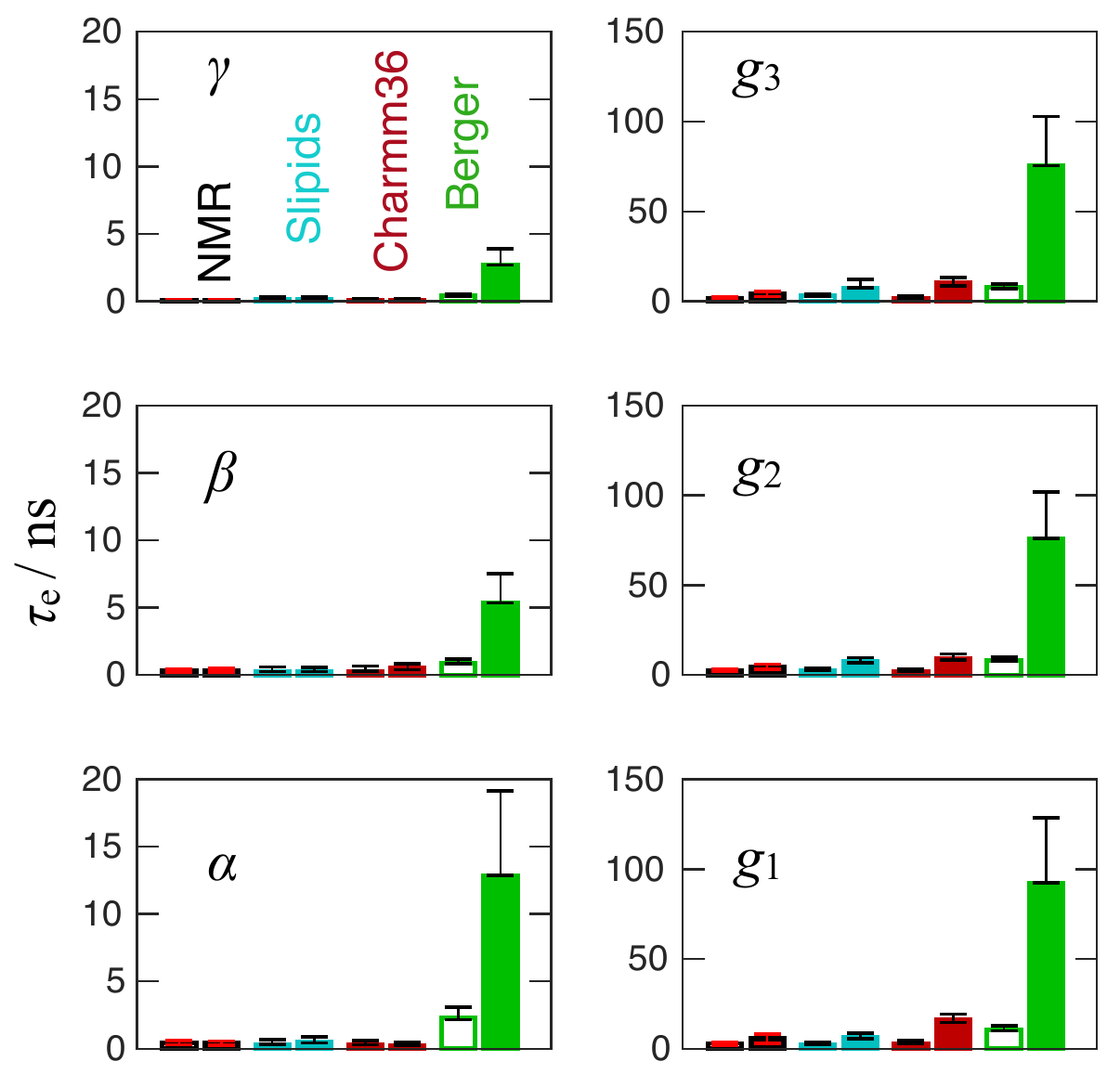}
    \caption{Same as Figure~\ref{fig_taue} but with proper scale for visualizing the Berger data. }
\label{fig_taue_Berger}
\end{figure*}

\begin{figure*}
  \centering
  \includegraphics[width=16cm]{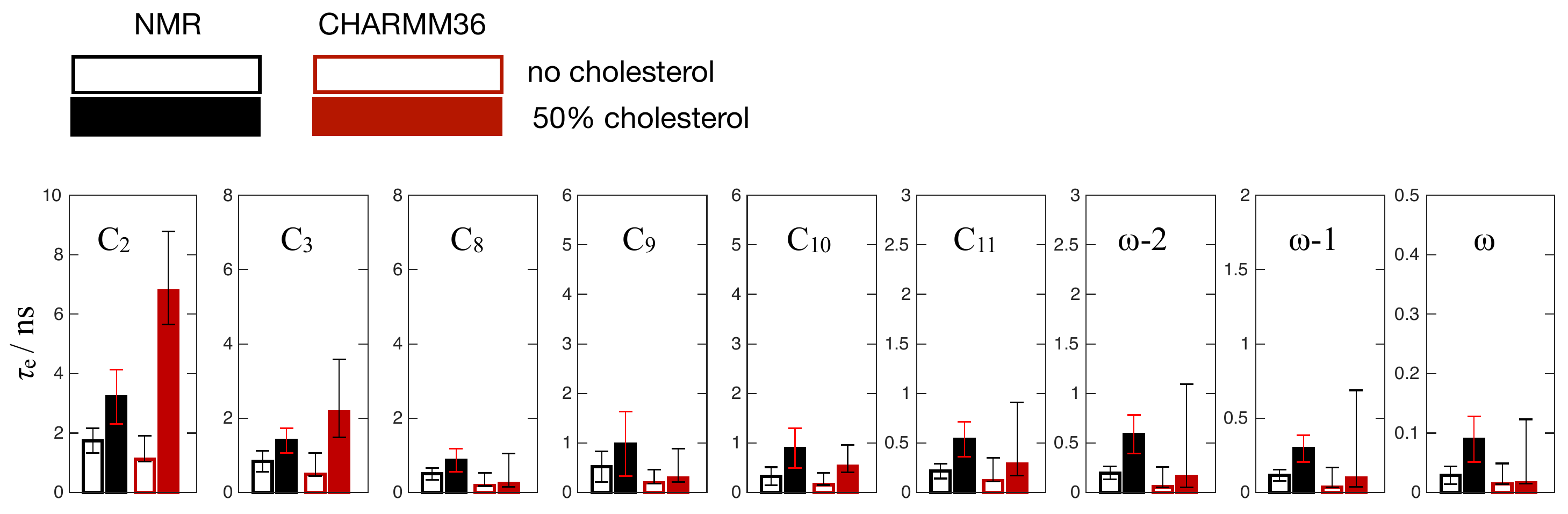}
  \caption{Effect of cholesterol on the effective correlation times, $\tau_{\rm{e}}$, of different segments in the acyl chains of POPC calculated experimentally and from lipid bilayer MD simulations with the CHARMM36 force field.
  }\label{taue_tails}
\end{figure*}

\begin{figure*}
  \centering
  \includegraphics[width=16cm]{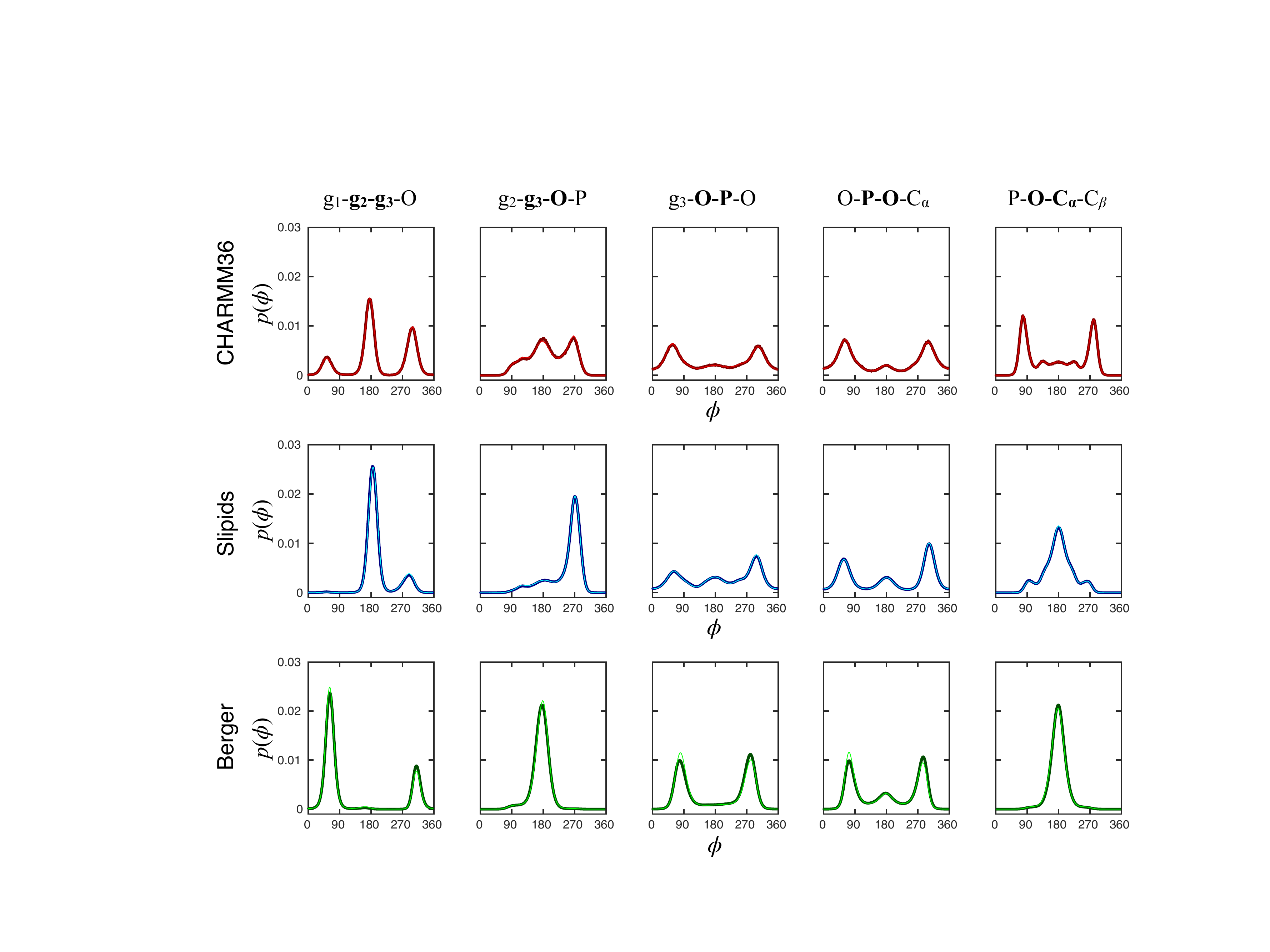}
  \caption{Dihedral angle distributions from the POPC (thick black lines) and POPC/cholesterol (thin colored lines) MD simulations using the CHARMM36 (top), Slipids (middle) and Berger (bottom) force-fields.  
  } \label{fig_dihedral_non_log_scale}
\end{figure*}

\begin{figure*}
  \centering
  \includegraphics[width=16cm]{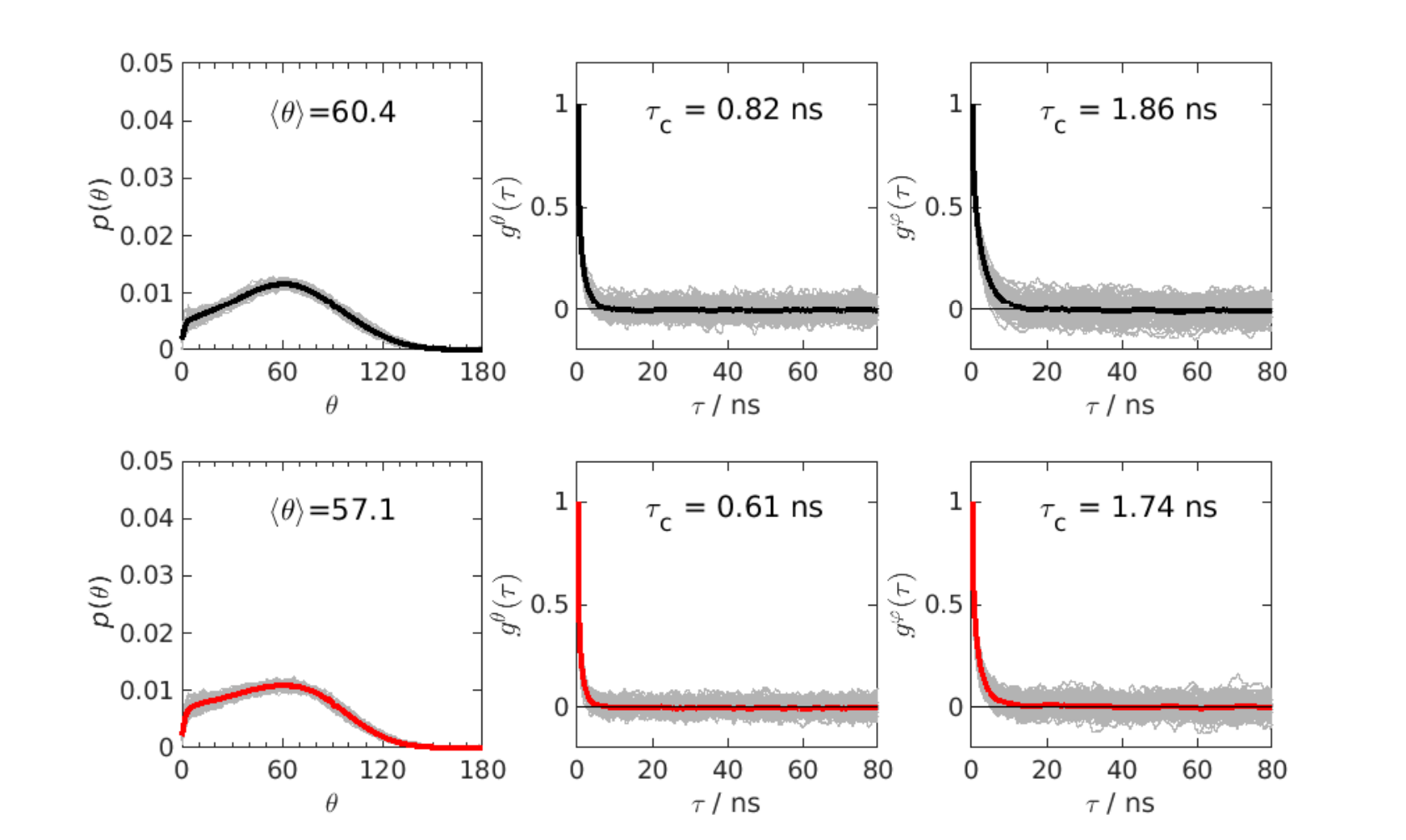}
  \caption{Orientation distribution, $p(\theta)$ (outermost left), and auto-correlation functions for the polar, $\theta$ (middle row), and azimuthal, $\varphi$ (outermost right), angles of vector P$\rightarrow$N from POPC (top) and POPC/cholesterol (bottom) CHARMM36 MD simulations. The grey lines represent data for each individual phospholipid molecule in the simulation. $\theta$ and $\varphi$ are the spherical angles in the laboratory coordinate frame defined by the simulation box axes. 
  } \label{PN_phi_theta}
\end{figure*}

\begin{figure*}
  \centering
  \includegraphics[width=16cm]{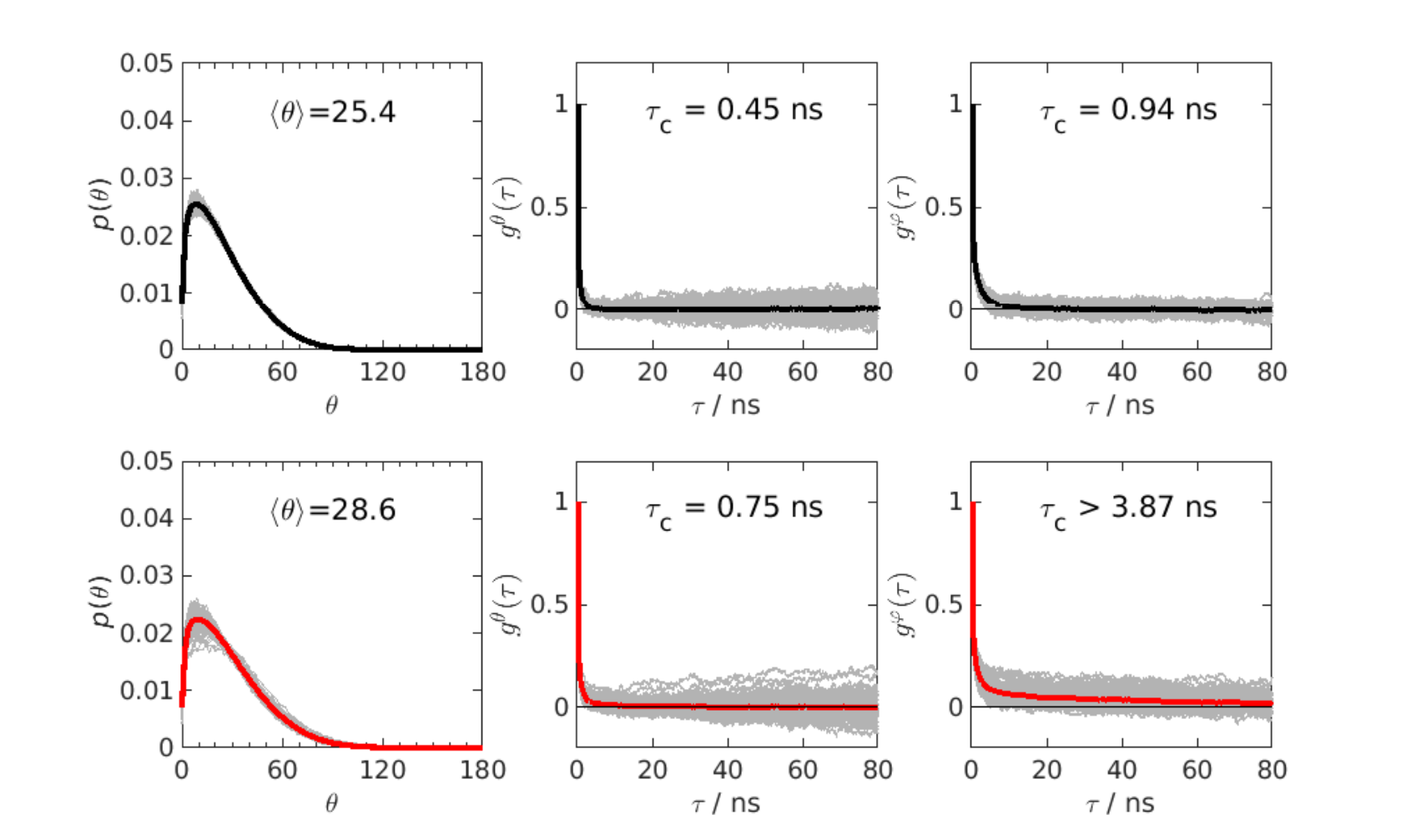}
  \caption{Orientation distribution, $p(\theta)$ (outermost left), and auto-correlation functions for the polar, $\theta$ (middle row), and azimuthal, $\varphi$ (outermost right), angles of vector ($g_3$-)O$\rightarrow$P  from POPC (top) and POPC/cholesterol (bottom) CHARMM36 MD simulations. The grey lines represent data for each individual phospholipid molecule in the simulation. $\theta$ and $\varphi$ are the spherical angles in the laboratory coordinate frame defined by the simulation box axes.
  } \label{O11_P_phi_theta}
\end{figure*}

\begin{figure*}
  \centering
  \includegraphics[width=16cm]{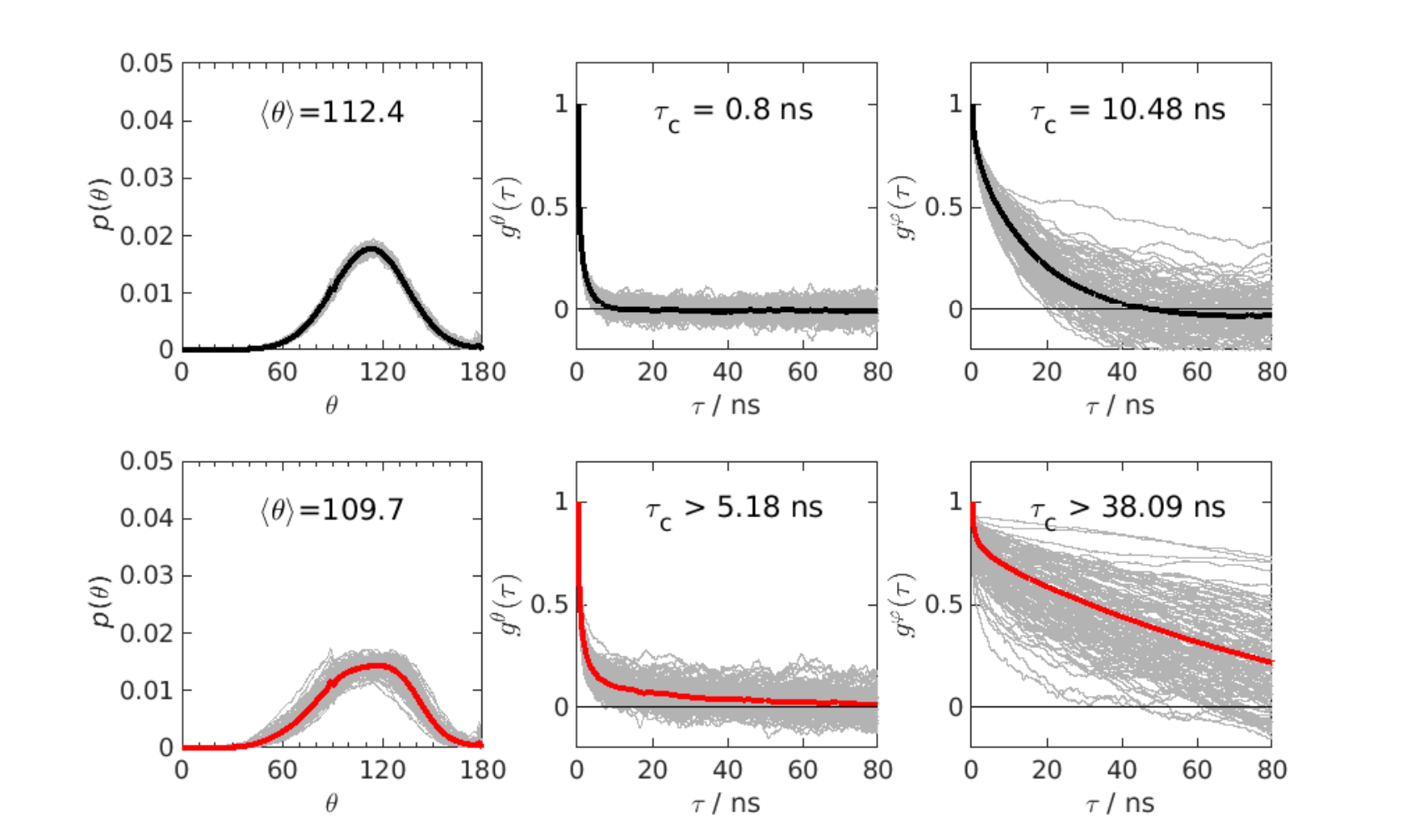}
  \caption{Orientation distribution, $p(\theta)$ (outermost left), and auto-correlation functions for the polar, $\theta$ (middle row), and azimuthal, $\varphi$ (outermost right), angles of vector ($g_1$-)O$\rightarrow$O(-$g_2)$ from POPC (top) and POPC/cholesterol (bottom) CHARMM36 MD simulations. The grey lines represent data for each individual phospholipid molecule in the simulation. $\theta$ and $\varphi$ are the spherical angles in the laboratory coordinate frame defined by the simulation box axes.
  } \label{O21_O31_phi_theta}
\end{figure*}

\begin{figure*}
  \centering
  \includegraphics[width=16cm]{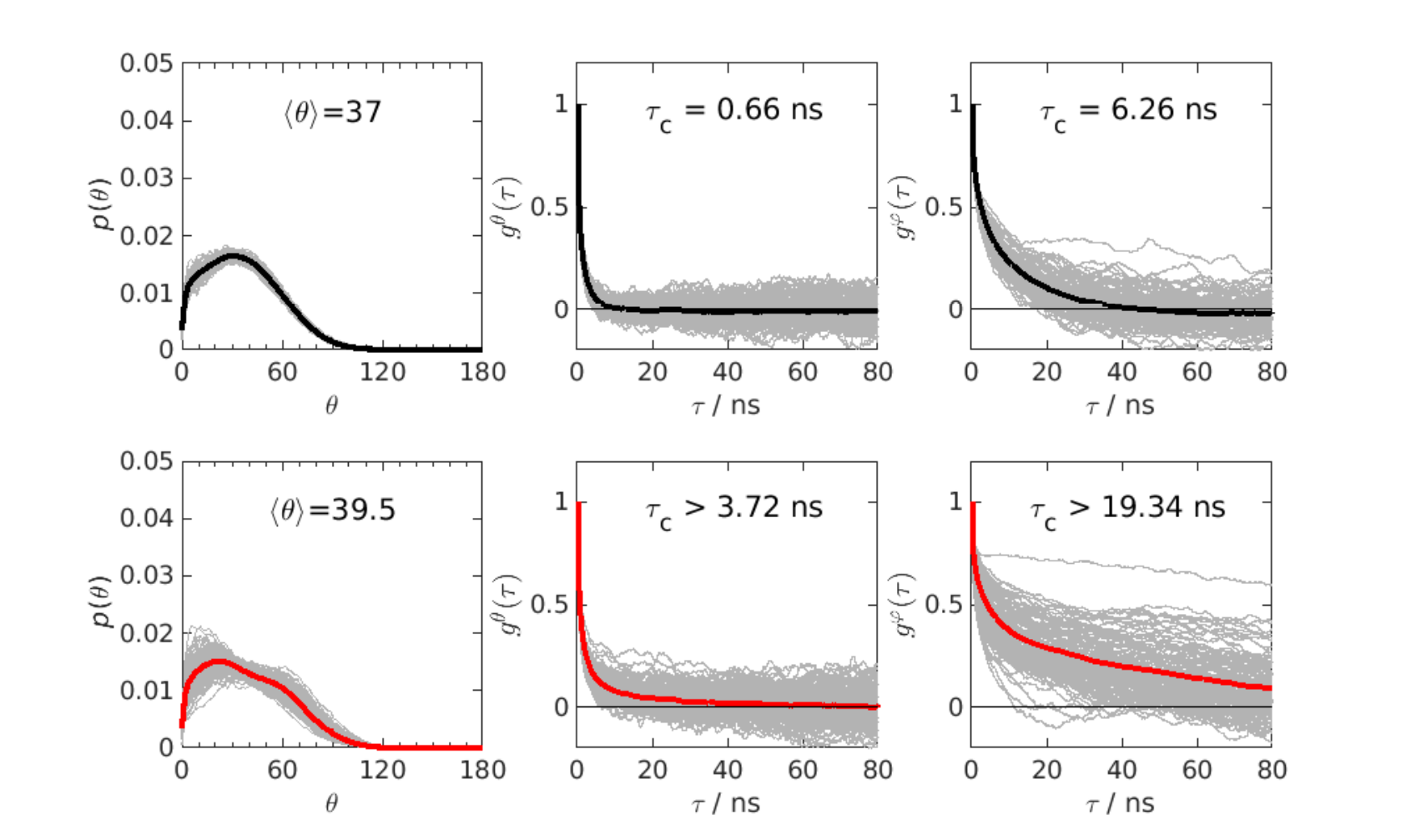}
  \caption{Orientation distribution, $p(\theta)$ (outermost left), and auto-correlation functions for the polar, $\theta$ (middle row), and azimuthal, $\varphi$ (outermost right), angles of vector $g_1\rightarrow g_3$ from POPC (top) and POPC/cholesterol (bottom) CHARMM36 MD simulations. The grey lines represent data for each individual phospholipid molecule in the simulation. $\theta$ and $\varphi$ are the spherical angles in the laboratory coordinate frame defined by the simulation box axes. 
  } \label{C3_C1_phi_theta}
\end{figure*}

\begin{figure*}
  \centering
  \includegraphics[width=16cm]{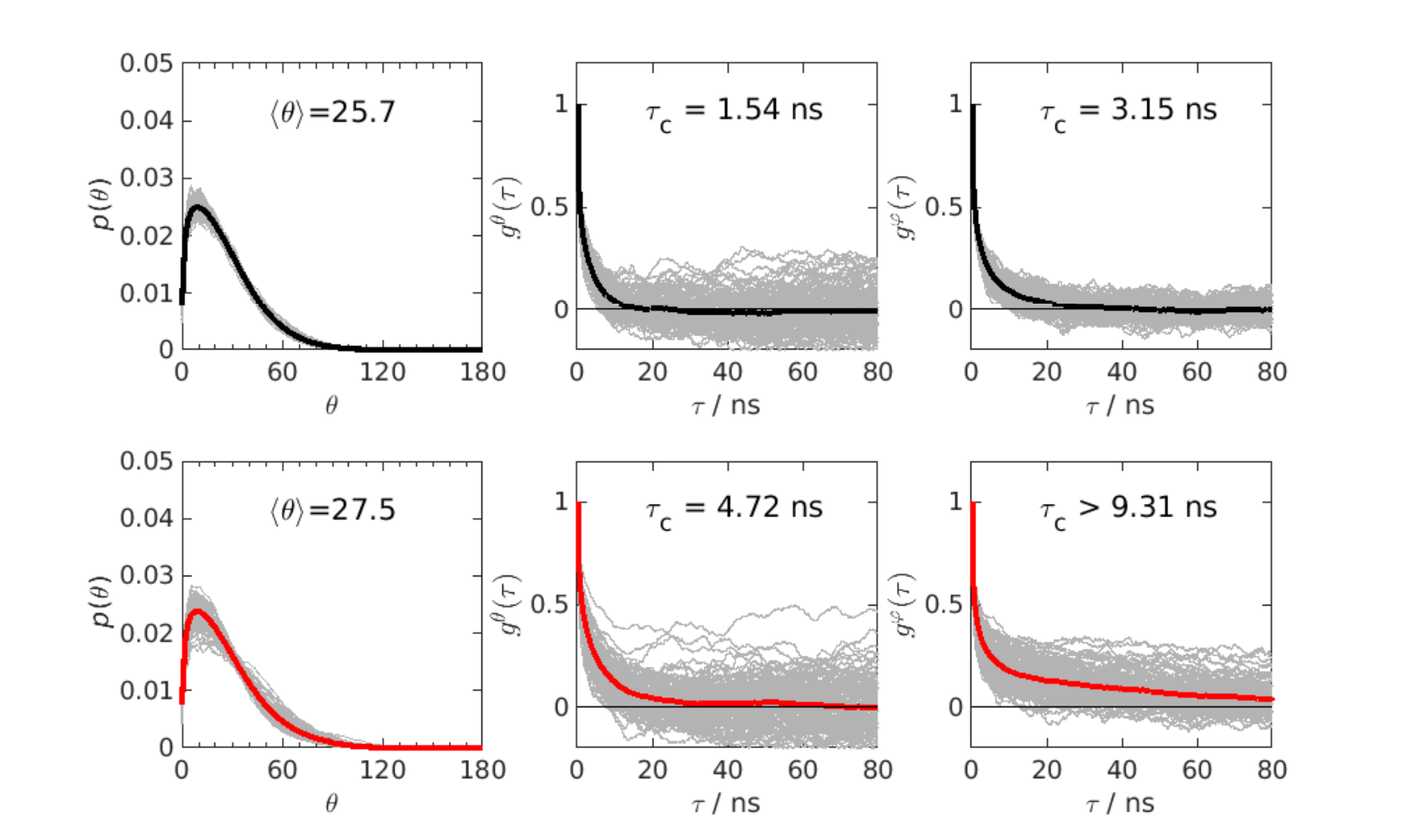}
  \caption{Orientation distribution, $p(\theta)$ (outermost left), and auto-correlation functions for the polar, $\theta$ (middle row), and azimuthal, $\varphi$ (outermost right), angles of vector $g_2\rightarrow g_3$ from POPC (top) and POPC/cholesterol (bottom) CHARMM36 MD simulations. The grey lines represent data for each individual phospholipid molecule in the simulation. $\theta$ and $\varphi$ are the spherical angles in the laboratory coordinate frame defined by the simulation box axes. 
  } \label{C2_C1_phi_theta}
\end{figure*}

\begin{figure*}
  \centering
  \includegraphics[width=16cm]{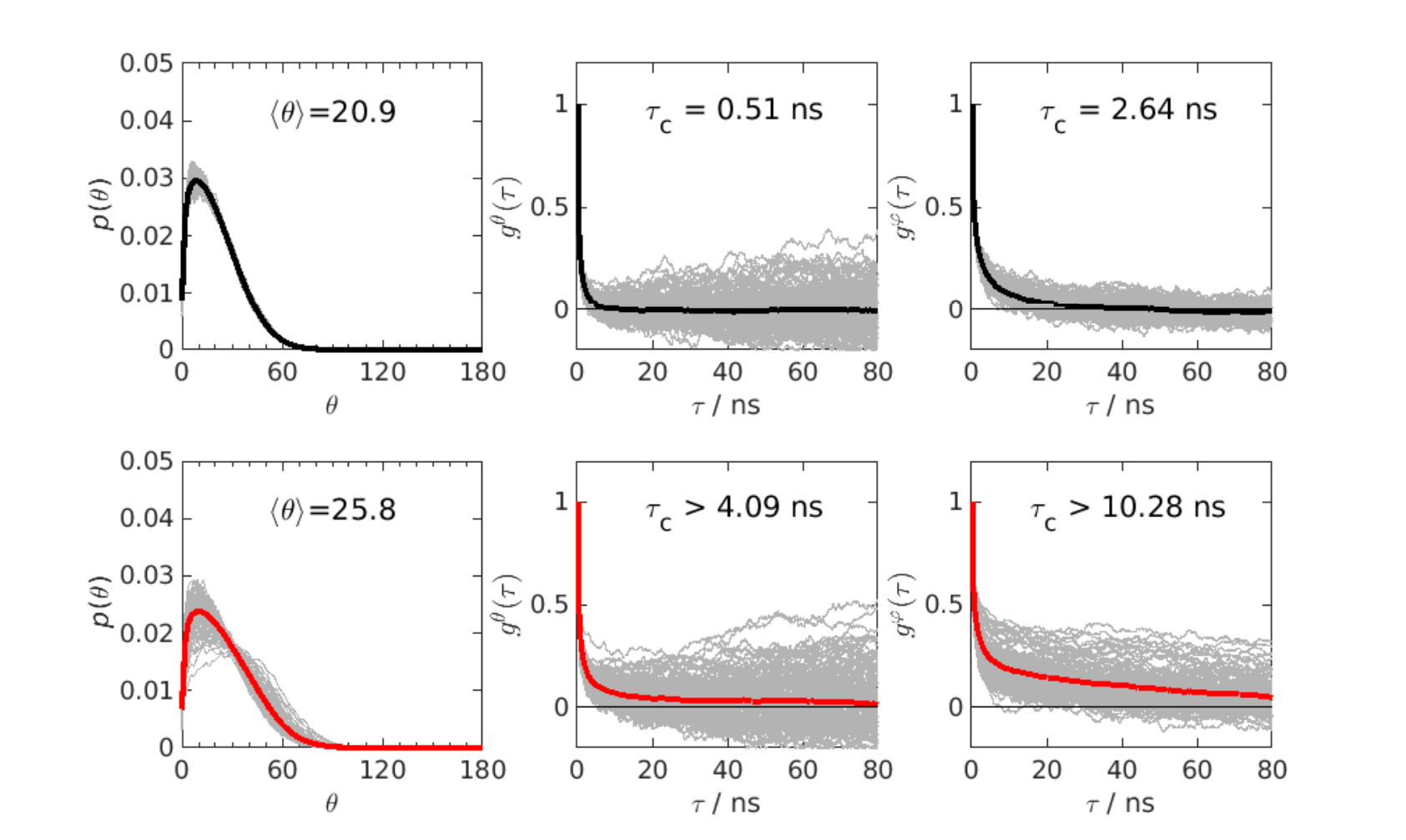}
  \caption{Orientation distribution, $p(\theta)$ (outermost left), and auto-correlation functions for the polar, $\theta$ (middle row), and azimuthal, $\varphi$ (outermost right), angles of vector $g_2$ $\rightarrow$P from POPC (top) and POPC/cholesterol (bottom) CHARMM36 MD simulations. The grey lines represent data for each individual phospholipid molecule in the simulation. $\theta$ and $\varphi$ are the spherical angles in the laboratory coordinate frame defined by the simulation box axes. 
  } \label{C2_P_phi_theta}
\end{figure*}

\begin{figure*}
  \centering
  \includegraphics[width=16cm]{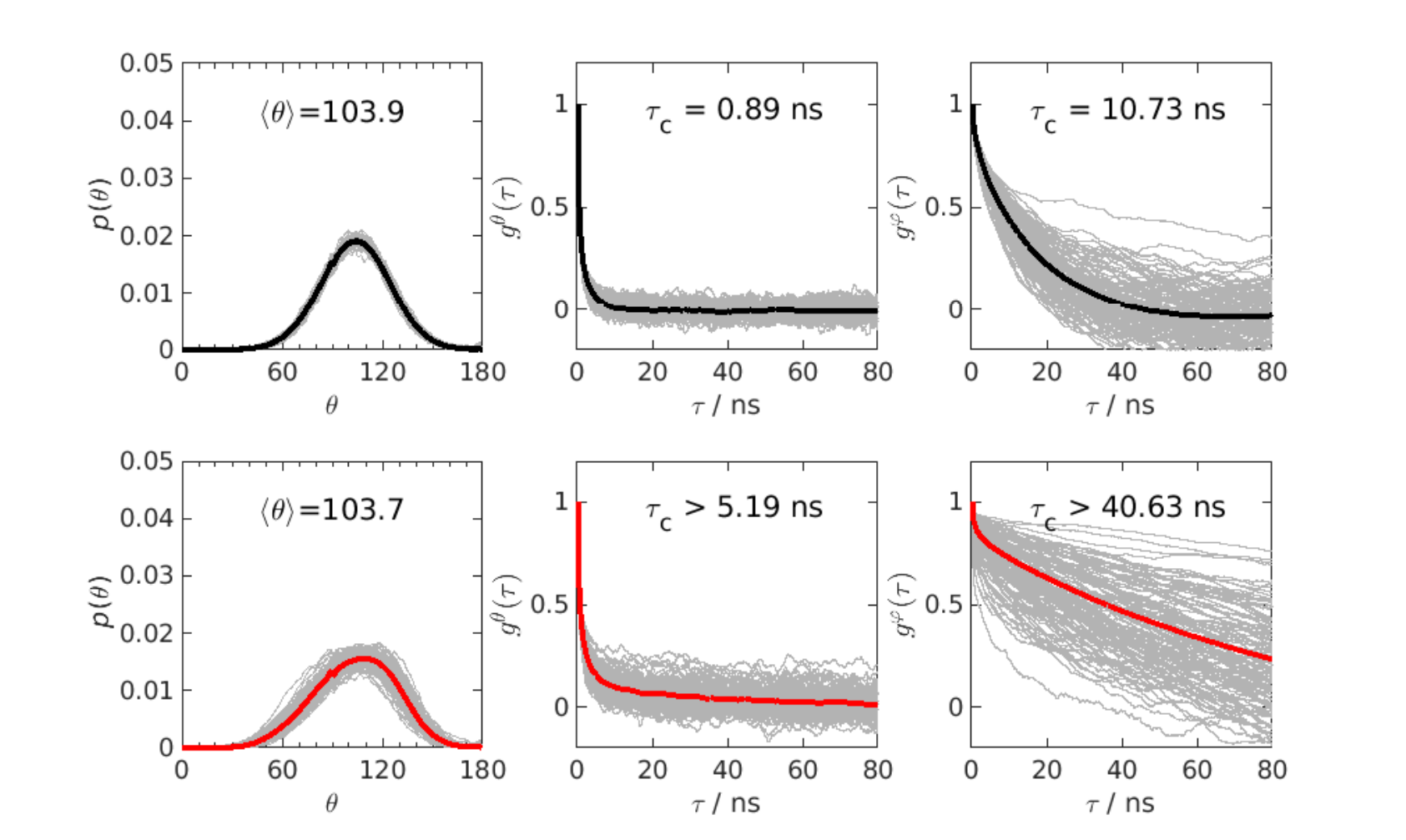}
  \caption{Orientation distribution, $p(\theta)$ (outermost left), and auto-correlation functions for the polar, $\theta$ (middle row), and azimuthal, $\varphi$ (outermost right), angles of vector C1({\it sn}-1)$\rightarrow$C1({\it sn}-2) from POPC (top) and POPC/cholesterol (bottom) CHARMM36 MD simulations. The grey lines represent data for each individual phospholipid molecule in the simulation. $\theta$ and $\varphi$ are the spherical angles in the laboratory coordinate frame defined by the simulation box axes. 
  } \label{C21_C31_phi_theta}
\end{figure*}

\end{document}